\documentclass[useAMS,usenatbib]{mn2e}

\pdfoutput=1
\usepackage{amssymb}
\usepackage{aas_macros}
\usepackage{natbib}
\usepackage{multirow}
\usepackage[draft]{hyperref}
\usepackage{color}
\usepackage{epsfig}
\usepackage{deluxetable}
\usepackage{multicol}
\usepackage{lscape}
\usepackage{pdflscape,pdfpages,subfig,float}
\usepackage{aecompl}

\title[Nuclear star clusters in the HST/WFPC2 archive]
{Nuclear\,star\,clusters\,in\,228\,spiral\,galaxies\,in\,the\,{\it HST/WFPC2} archive: catalogue and comparison to other stellar systems
}
\author[I. Y. Georgiev and T. B\"oker]{Iskren Y. Georgiev$^{1}$\thanks{E-mail:
iskren.georgiev@esa.int; iskren.y.g@gmail.com
} and Torsten B\"oker$^{2}$\\
$^{1}$European Space Agency - ESTEC, Keplerlaan 1, 2201 AG Noordwijk, Netherlands\\
$^{2}$European Space Agency - Space Telescope Science Institute, 3700 San Martin Drive, Baltimore, MD 21218, USA}

\begin{document}

\date{Accepted 2014 April 17; in original form 2014 January 27}

\pagerange{\pageref{firstpage}--\pageref{lastpage}} \pubyear{2014}

\maketitle

\label{firstpage}

\begin{abstract}

We present a catalogue of photometric and structural properties of 228 nuclear star clusters (NSCs) in nearby late-type disk galaxies. These new measurements are derived from a homogeneous analysis of all suitable WFPC2 images in the {\it HST} archive. The luminosity and size of each NSC is derived from an iterative PSF-fitting technique, which adapts the fitting area to the effective radius ($r_{\rm eff}$) of the NSC, and uses a WFPC2-specific PSF mo\-del tailored to the position of each NSC on the detector.

The luminosities of NSCs are $\leq\!10^8L_{\rm V,\odot}$, and their integrated optical colours suggest a wide spread in age. We confirm that most NSCs have sizes similar to Globular Clusters (GCs), but find that the largest and brightest NSCs occupy the regime between Ultra Compact Dwarf (UCD) and the nuclei of early-type galaxies in the size-luminosity plane. The overlap in size, mass, and colour between the different incarnations of compact stellar systems provides a support for the notion that at least some UCDs and the most massive Galactic GCs, may be remnant nuclei of disrupted disk galaxies.

We find tentative evidence for the NSCs' $r_{\rm eff}$ to be smaller when measured in bluer filters, and discuss possible implications of this result. We also highlight a few examples of complex nuclear morphologies, including double nuc\-lei, extended stellar structures, and nuclear $F606W$\,excess from either recent (circum-)nuclear star formation and/or a weak AGN. Such examples may serve as case studies for ongoing NSC evolution via the two main suggested mechanisms, namely cluster merging and {\it in situ} star formation. 

\end{abstract}


\section{Introduction}
Driven mostly by advances in the spatial resolution of modern telescopes over the last decades, it has now become firmly established that nuclear star clusters (NSCs) are an important morphological component of all types of galaxies \cite[e.g.][]{Phillips96, Carollo98, Boeker02,Boeker04,Cote06,Georgiev09b,Turner12}. 

The connection between the formation and evolution of NSCs and their host galaxies is a much-discussed topic of modern astrophysics. In particular, it is an open question whether NSCs are an essential ingredient for (or an intermediate step towards) the formation of a supermassive black hole (SMBH) in the galaxy nucleus \citep{Neumayer&Walcher12}. This question has been brought into focus by the realization that in many galaxies, both NSC and SMBH co-exist \citep{Seth08,Graham&Spitler09}, and that the few known SMBHs in bulge-less disks all reside in NSCs \citep{Filippenko&Sargent89, Shields08, Satyapal08, Satyapal09, Barth09, Secrest12}. 

The debate on the interplay between NSCs and SMBHs has been fuelled further by the finding that both types of "central massive object" (CMO) appear to grow in a way that is correlated with the growth of their host galaxies. This correlation has been induced from a number of so-called scaling relations, which demonstrate the dependence of CMO mass on various properties of the host galaxy. More specifically, the mass of both SMBHs \cite[e.g.][]{Ferrarese&Merritt00, Gebhardt00,Haering&Rix04} and NSCs \cite[e.g.][]{Wehner&Harris06, Rossa06, Ferrarese06} appears to correlate with the mass of the host galaxy bulge \cite[see][for a recent summary of this topic]{Scott&Graham13}. The most promising way to investigate the driving mechanism(s) behind these scaling relations is perhaps the study of late-type disk galaxies which are believed to be the most "pristine" galaxies which have not (yet) experienced any significant build-up of either bulge or CMO, and should therefore be well-suited  to investigate the early stages of their (co)evolution. 

Understanding the origin and evolution of NSCs may also shed light on the nature of other massive compact stellar systems such as Globular Clusters (GCs) and Ultra Compact Dwarf galaxies (UCDs). There are numerous suggested scenarios for the origin of UCDs, 
including them being the extreme end of the GC luminosity function \cite[e.g.][]{Drinkwater00,Mieske02,Mieske12}, the end product of star cluster merging \cite[e.g][]{Kroupa98,Fellhauer&Kroupa02a,Kissler-Patig06,Bruens11}, the former nuclei of now dissolved galaxies \cite[e.g.][]{Bekki01,Bekki&Freeman03,Ideta&Makino04, Pfeffer&Baumgardt13}, or a combination of these mechanisms \cite[e.g.][]{Mieske06,Hilker09,DaRocha11,Brodie11,Norris&Kannappan11}. 

In particular, expanding the sample of NSCs with well-characterized sizes and stellar populations is needed to provide empirical constraints on the ``stripped dwarf galaxy'' scenario. The latter has recently received observational support from an overlap in the properties of UCDs and dwarf galaxy nuclei, which appear to show similar trends in their internal velocity dispersions \cite[e.g.][]{Drinkwater03,Chilingarian11,Frank11}, size-luminosity and color-magnitude relations \cite[e.g.][]{Cote06,Rejkuba07,Evstigneeva08,Taylor10}, their luminosity-weighted integrated ages and metallicities \cite[e.g.][]{Paudel10,Paudel11,Francis12,Madrid13}, and dynamical mass-to-light ratios \citep{Hasegan05,Hilker07,Mieske08,Taylor10} which seem to suggest unusual stellar mass functions \cite[e.g.][]{Mieske&Kroupa08,Dabringhausen09,Dabringhausen12,Marks12,Bekki13} or the presence of dark matter, most likely in the form of a SMBH \citep{Mieske13}. All these observations hint at a close connection between UCDs (and high-mass GCs) and NSCs, which should be further tested by comparison to NSC covering as wide a range in size and mass as possible. Further connections can be provided from utilizing the high spatial resolution of HST, to enable the investigation of internal spatial variations of the stellar populations of such systems \cite[e.g.][]{Kundu&Whitmore98,Larsen01,Strader12,Sippel12,Wang&Ma13,Puzia14}.

Given that the typical sizes of NSCs fall into the range between a few pc and a few tens of pc \citep{Boeker04,Cote06,Turner12}, measuring their effective radii (and accurately separating their light from the surrounding, often complex, galaxy structure) requires HST resolution in all but the closest galaxies.
We have therefore explored the HST/WFPC2 Legacy archive to analyse all available exposures of spiral galaxies within $\leq40$\,Mpc, 
and to derive the structural and photometric properties of the identified NSCs. Taking advantage of the accurate instrument knowledge gained following nearly 20 years of WFPC2 observations, our work expands on previous studies by i) significantly increasing the number of NSCs with accurate size and flux measurements, ii) improving the accuracy of previous photometric measurements by using updated PSF-fitting techniques, and iii) increasing (by a factor of three) the number of NSCs within an expanded morphological range of late-type spiral galaxies. This will allow to study evolutionary trends with the Hubble type of the host galaxy.

Our work is organized as follows. In Section\,\ref{Sect:Data-Reduction-Analysis} we describe the galaxy sample and NSC identification (Sect.\,\ref{Sect:Gal-samp}). Image processing and combination is discussed in Section\,\ref{Sect:Image processing and combination}. Section\,\ref{Sect: NSCs} details the PSF-fitting techniques to derive the NSCs' structural parameters (Sect.\,\ref{Sect: Measuring NSC sizes}) and photometry (Sect.\,\ref{Sect: NSC photometry}). The limitations and uncertainties of the measured sizes, ellipticities and photometry are discussed in Section\,\ref{Sect: Uncertainties} and comparison with earlier work is performed in Section\,\ref{Sect:Comparison to Previous Studies}. Analysis of the general properties of the NSCs' size and luminosity distributions are presented in Sections\,\ref{Sect: Size Distribution} and \ref{Sect: NSC stellar pops}. We discuss the implications for the formation and evolution of massive compact stellar systems (\S\,\ref{Sect: NSCs_GC_UCDs}), the growth of NSCs (\S\,\ref{Sect: multiple NSCs}), and their coexistence with weak AGNs (\S\,\ref{Sect: AGNs}). Finally, we summarize our results in \S\,\ref{Sect:summary}.

%%%%%%%%%%%%%%%%%%%%%%%
\section{Data, Reduction, and Analysis}\label{Sect:Data-Reduction-Analysis}
%%%%%%%%%%%%%%%%%%%%%%%

\subsection{Galaxy sample and NSC identification}\label{Sect:Gal-samp}
We searched the HST/WFPC2 archive for all exposures of galaxies with late Hubble type ($t\!\geq\!3.5$) to avoid the most luminous bulges, an inclination of $i\!\leq\!88^\circ$ to avoid edge-on galaxies, and distances of $\leq\!40$\,Mpc, $(m-M)\!\lesssim\!33$\,mag to be able to reliably measure the size of the NSC (see \S\,\ref{Sect: Measuring NSC sizes} and Fig.\,\ref{fig: reff_dist} for a more detailed discussion of the resolution limit).
Because the presence of a strong AGN will complicate or even prevent the NSC characterization, we excluded all strong AGNs from the search, based on their {\tt agnclas} parameter in HyperLEDA. However, due to technical issues with searching and retrieving data from the archive, a few weak AGNs ended up in our sample, and were processed through our analysis pipeline. We nevertheless decided to use the measured NSC properties of these galaxies for a comparison to those of quiescent nuclei, and to check whether the presence of a weak AGN can be deduced from this comparison. This is discussed further in Section\,\ref{Sect: AGNs}.

We first created a list of all galaxies meeting the above criteria by searching the HyperLeda database\footnote{http://leda.univ-lyon1.fr/leda/fullsql.html} \citep{Paturel03}. 
We then used the coordinates of all galaxies returned by HyperLeda to query the HST archive for available Wide Field and Planetary Camera 2 (WFPC2) imaging within a 2\arcmin\ search radius. The search was limited to the well calibrated broad-band filters $F300W$, $F336W$, $F380W$, $F439W$, $F450W$, $F555W$, $F606W$, $F675W$, and $F814W$. For the archive query, we used the ESAC interface\footnote{http://archives.esac.esa.int/hst/} 
to the HST archive. This interface provides a preview image for each exposure, which we used for an initial examination in order to reject exposures which do not contain the galaxy nucleus within the WFPC2 field of view, or galaxies that are misclassified in either Hubble-type and/or inclination. The resulting number of galaxies with at least one suitable WFPC2 exposure in the HST archive is 323. In total, we retrieved data from 47 different GO and SNAP programs\footnote{HST/WFPC2 data from GO and SNAP programs: 5375, 5381, 5396, 5397, 5411, 5415, 5427, 5446, 5479, 5962, 5999, 6231, 6232, 6355, 6359, 6367, 6423, 6431, 6483, 6713, 6738, 6833, 6888, 7450, 8192, 8199, 8234, 8255, 8597, 8599, 8601, 8632, 8645, 9042, 9124, 9720, 10803, 10829, 10877, 10889, 10905, 11128, 11171, 11227, 11603, 11966, 11987}. 

The identification of the galaxy's nucleus and any NSC in its center is a relatively straightforward task for early-type (spheroidal) galaxies. In late-type galaxies, however, this task is often complicated by ongoing star formation in the vicinity of the nucleus with its many manifestations: bright disks/rings, multiple star-forming complexes, bars (often off-centred), dust lanes, etc. We therefore visually inspected all downloaded exposures in order to identify those with an unambiguous NSC. 

\begin{figure}
\includegraphics[scale=0.31]{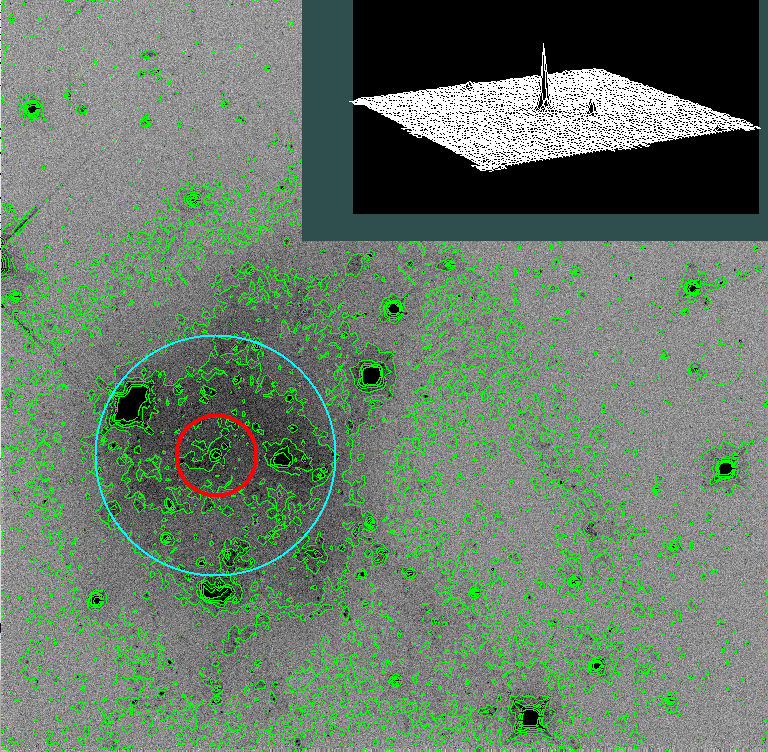}
\caption{WFPC2/WF3 F606W image of ESO271-G005, an example for a very late-type ($t\!=\!9$, Sm) galaxy. The NSC is the brightest source near the photocenter, as demonstrated by the general agreement between the outer iso-intensity contour and the two circular apertures centred on the NSC. The inner and outer circles around the NSC have radii of 40 and 120 pixels, respectively. The size of the mesh plot in the upper right is $80\!\times\!80$ pixels. The outermost iso-intensity contour marks $10\!\times$ the background level. Other bright sources in the image are star forming regions and foreground stars.  }
\label{fig:irregular NSC}
\end{figure}

There are 27 galaxies in our sample with a morphological type of Sm or Irr (i.e. $t\geq9$). It is often difficult to identify the galaxy nucleus in such galaxies with a perturbed morphology. For these cases, we identified the brightest star cluster (which often is the only one) near the photometric center as the galaxy's NSC. In order to check whether its location plausibly defines the galaxy nucleus, we centred a large-diameter circular aperture on it, and compared this to the outer iso-intensity contours. An example for this is shown in Figure\,\ref{fig:irregular NSC}.

During the inspection, 95 galaxies were removed from the catalogue as unsuitable for reliable NSC fitting. There are a variety of reasons for the rejection. Some are related to the exposure quality (e.g. poor signal-to-noise ratio, saturation, or the nucleus falling outside or too close to the detector edge), while others are intrinsic to the galaxy (e.g. the genuine absence of any prominent cluster close to the photocenter, the presence of multiple clusters of comparable luminosity, or a generally complex structure of the nuclear region). All of these prevent a reliable fit of the NSC structure and photometry with our PSF fitting techniques. For completeness, Table\,\ref{Table:Excluded galaxies} lists these 95 galaxies and their primary reason for rejection. Some examples for such complex structures, including double nuclei or extended nuclear disks  are presented in Sect.\ref{Sect: multiple NSCs}). 

\begin{figure}
\epsfig{file=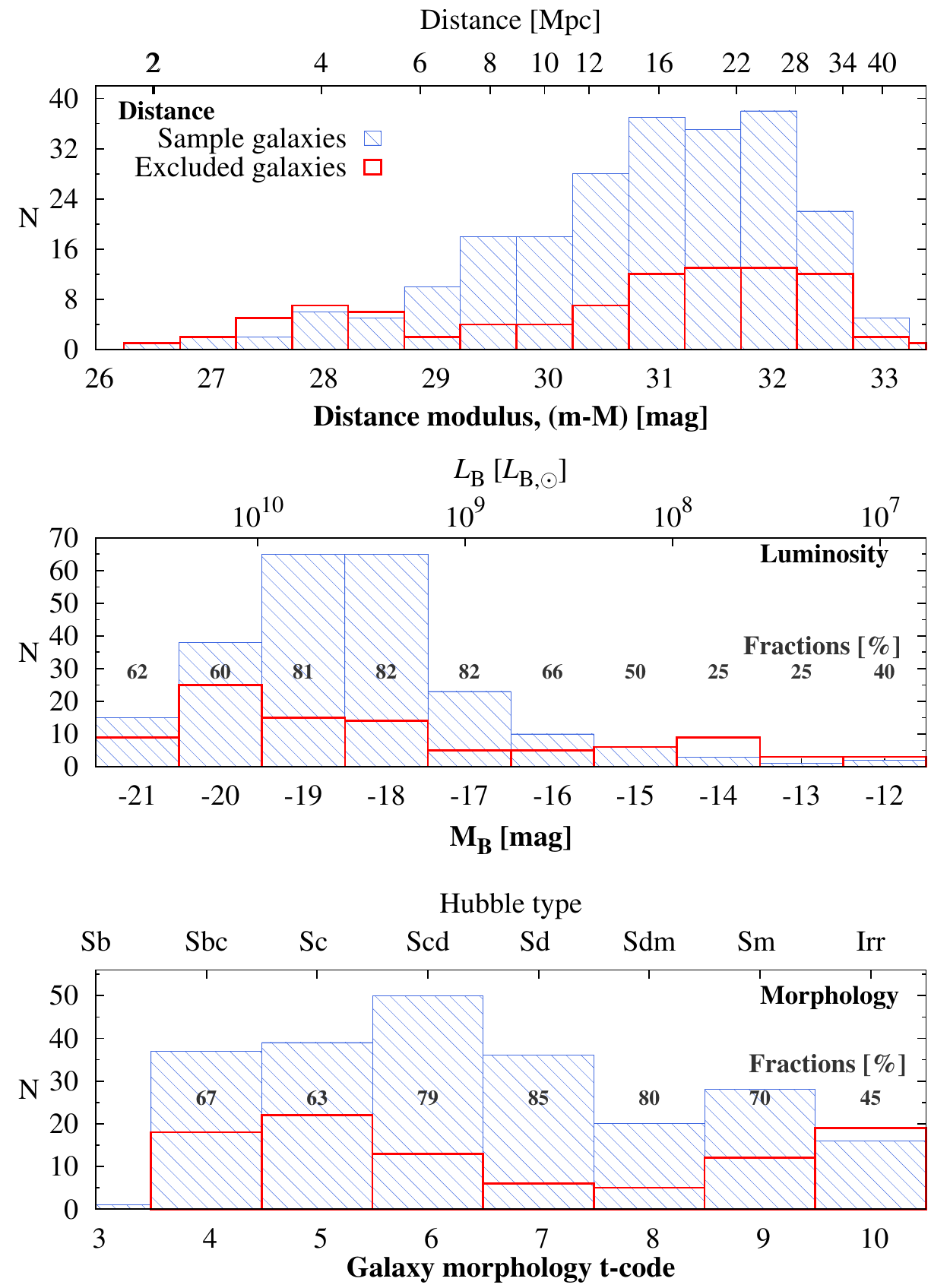, width=0.5\textwidth, bb=20 10 360 490}
\caption{Histograms of the main sample properties, i.e. distance (top), luminosity (middle), and Hubble type (bottom). Separate histograms are plotted for the NSC catalog proper (hashed), and for the sample of rejected galaxies (open), as discussed in \S\,\ref{Sect:Gal-samp}.
\label{fig:Galaxy sample}}
\end{figure}

The final NSC catalogue discussed in this paper therefore contains 228 objects. The main properties of the NSC host galaxy sample collected from the HyperLeda and NED databases, are summarized in Table\,\ref{Table:Galaxy sample}, and illustrated in Figure\,\ref{fig:Galaxy sample} with distributions of distance,
luminosity and morphological type. In Figure\,\ref{fig:Galaxy sample}  we also show the respective distributions of the rejected galaxy sample, in order to enable a discussion of the nucleation fraction of late-type galaxies in the next section.

%

%%%%%%%%%%%%%%%%%%%%%%%
\subsection{Nucleation Fraction}
%%%%%%%%%%%%%%%%%%%%%%%
%
For each bin in the lower two panels of Figure\,\ref{fig:Galaxy sample}, the overplotted numbers indicate the fraction (in \%) of galaxies with a well-fitted NSC over the total number of galaxies in that bin. It is important to point out that these numbers can only serve as lower limits to the true nucleation fraction, given the variety of possible reasons for the rejection of galaxies. For example, a galaxy that was rejected because it falls into the Extended/Complex or Multiple categories may still harbour a genuine NSC which, however, cannot be identified easily, or measured reliably. 

With these caveats in mind, the numbers in Figure\,\ref{fig:Galaxy sample} show that on average 80\% of all late-type galaxies we retrieved from the HST/WFPC2 archive harbour a well defined nuclear star cluster. This confirms the results of earlier studies \citep{Boeker02} which also find that at least 80\% of the late-type disk galaxies ($6\!\leq\!t\!\leq\!9$) harbor an unambiguous NSC. There is some evidence for a decreasing nucleation fraction in earlier Hubble types ($t < 6$), as well as in irregular galaxies ($t=10$). While the lower nucleation fraction in Irregulars suggested by Figure\,\ref{fig:Galaxy sample} may be affected by small-number statistics, the numbers are in agreement with those found for low-luminosity dIrrs ($M_V>-17$\,mag) of about 10\%  \cite[7/68 galaxies in ][]{Georgiev09b}.

%%%%%%%%%%%%%%%%%%%%%%%
\subsection{Image processing and combination}\label{Sect:Image processing and combination}
%%%%%%%%%%%%%%%%%%%%%%%

All WFPC2 exposures retrieved from the Hubble archive are processed with the WFPC2 {\tt calwp2}\footnote{\href{http://www.stsci.edu/hst/wfpc2/wfpc2_reproc.html}{http://www.stsci.edu/hst/wfpc2/wfpc2\_reproc.html}} instrument pipeline. It uses the latest calibration reference files to correct for bias, dark current, detector response variations (flat-fielding), as well as various electronic artefacts (e.g. reduced 34th row size and WF4 gain degradation). We then corrected the downloaded exposures for bad pixels using the latest masks that were retrieved together with the science data. In addition, we removed cosmic rays (CRs) hits with \mbox{\sc lacosmic}, a Laplacian kernel identification algorithm described and made available as an IRAF\footnote{IRAF is distributed by the National Optical Astronomy Observatories, which are operated by the Association of Universities for Research in Astronomy, Inc., under cooperative agreement with the National Science Foundation.} procedure\footnote{\href{http://www.astro.yale.edu/dokkum/lacosmic}{http://www.astro.yale.edu/dokkum/lacosmic}} by \cite{vanDokkum01}. We carefully tested the {\sc lacosmic} parameters\footnote{We find the following main {\sc lacosmic} parameters to work well in removing the majority of CRs: {\tt sigfrac=0.2, objlim=5, niter=4}} to remove only CRs while avoiding the tips of bright stars and/or compact star clusters. A final image combination of multiple exposures per filter (if available) helped to remove any remaining CRs using sigma and percentile pixel clipping algorithms. 

For each WFPC2 filter, we registered and combined the individual exposures  using a custom-written IRAF wrapper procedure to combine and efficiently automate individual steps with IRAF procedures. Using an initial guess from selected stars in the reference image, the code identifies high-$S/N$ stars present in all exposures, evaluates and applies the subpixel shifts (with a drizzle re-sampling factor chosen to be 0.6), applies corrections for field rotation and distortion (with {\sc geomap, geotran}), and combines the registered images ({\sc imcombine}) after scaling them for exposure time and correcting for any remaining zero level offsets. The zero levels are estimated from a specified $30\times30$ pixels statistic section region that is selected to be free of contaminating sources, defined as a header keyword added during an earlier preparatory step. The achieved accuracy in the image registration is about 0.08 pixels (RMS) for exposures taken with a dither pattern, and better for observations obtained with a non-dithered (CR-split) strategy, which is the case for the majority of the data. A few long exposures with integration times ($>\!1000$\,seconds) and/or dithered exposures were found to have a fine field rotation of up to ($0.08\degr$) which was corrected as well.

%%%%%%%%%%%%%%%%%%%%%%%
\section{Analysis of Nuclear Star Clusters}\label{Sect: NSCs}
%%%%%%%%%%%%%%%%%%%%%%%

%%%%%%%%%%%%%%%%%%%%%%%
\subsection{Measuring NSC sizes}\label{Sect: Measuring NSC sizes}
%%%%%%%%%%%%%%%%%%%%%%%
%
The present-day structure of a compact stellar system bears witness to its evolutionary past, i.e. the combined effects of the internal dynamical processes and the gravitational potential to which the system as a whole is subjected. Because of the limited spatial resolution, in observations of extragalactic star clusters, it is generally hard, if not impossible, to measure the shape of the surface brightness profile (and hence mass density profile) with sufficient accuracy to distinguish between various predictions and/or models. Therefore, the light profile of extragalactic clusters is typically compared to that of the instrument Point Spread Function (PSF) convolved with an analytical function which is known to represent well the structure of resolved Galactic globular clusters \cite[surface-brightness profile, concentration, core, effective, tidal radii, e.g.][]{King66,EFF87}. Assuming a precise knowledge of the instrumental PSF, this approach can yield a reliable measurement of the effective radius $r_{\rm eff}$, i.e. the radius that contains half the cluster light.

The HST PSF is very well characterized: the {\sc TinyTim}\footnote{\href{http://www.stsci.edu/hst/observatory/focus/TinyTim}{http://www.stsci.edu/hst/observatory/focus/TinyTim}} software package allows to create a PSF model corrected for a multitude of factors that influence the PSF shape such as the precise HST focus position (a.k.a. breathing), the instrument used, detector chip, position within the chip, filter, the object's spectral type, charge transfer effects, etc. \citep{Krist&Hook11}. All of these factors are properly accounted for when
constructing the PSF model for each exposure.

Because NSCs in late-type galaxies are known to contain a mix of stellar populations \cite[e.g.][]{Walcher06,Seth06,Seth10}, we chose an intermediate-type spectral energy distribution (SED) of an F8V-type star ($V-I=0.68$\,mag) for the generation of the {\sc TinyTim} PSFs which are then oversampled by a factor of ten\footnote{The ten times oversampled PSF is convolved with the charge diffusion kernel, provided as a separate file by {\sc TinyTim}, during the PSF fitting with {\sc ishape}, which restores the charge diffusion blurring.}, and tailored to the NSC position on the respective WFPC2 detector. As discussed in Sect.\,\ref{Sect: Uncertainties}, the impact of this particular choice of SED on the derived NSC properties is small.

To quantitatively compare NSCs in our sample to the PSF shape, we used the {\sc ishape} procedure in the {\sc baolab} software package\footnote{\href{http://baolab.astroduo.org}{http://baolab.astroduo.org}} \citep{Larsen99}. {\sc ishape} measures the size of a compact source via an algorithm that minimizes the $\chi^2$ difference between the observed light profile and that of a model cluster. The latter is generated by convolving the instrumental PSF with a choice of analytical models available in {\sc ishape}. More specifically, we use various versions of a tidally truncated isothermal sphere \cite[or King-profiles,][]{King62,King66} with concentration indices ($r_{\rm tidal}/r_{\rm core}$) of 5, 15, 30, and 100, as well as power law profiles \cite[EFF,][]{EFF87} for two indices of 1.5 and 2.5.
The {\sc ishape} model best describing the data (i.e. having the smallest $\chi^2$ residuals) is then used to derive both the effective radius as well as the flux of the NSC.

For high signal-to-noise data with $S/N\!>\!30$, {\sc ishape} can provide a reliable measurement of $r_{\rm eff}$ for intrinsic sizes as small as 10\% of the PSF \citep{Larsen99}, i.e. 0.2 pixels on the WFPC2/PC detector. This implies that for a galaxy at 30\,Mpc distance, effective radii as small as $r_{\rm eff}\simeq3$\,pc can be reliably measured 
(see discussion in \S\,\ref{Sect: Uncertainties}).

To facilitate an automated fitting process of NSCs in the large and heterogeneous galaxy sample, we developed a wrapper procedure in the IRAF environment, which ports {\sc ishape} and all its parameters. The procedure reads the image name and NSC position from an input list created during the visual identification described in Sec. \ref{Sect:Gal-samp}. It reads from the image header all necessary information such as WFPC2 detector chip, filter, distance to galaxy, etc.. 

A robust measurement of the NSC sizes (as well as fluxes) requires that the {\sc ishape} fitting radius extends to about three times the object's FWHM. Our procedure therefore performs an iterative adjustment of the {\sc ishape} fitting radius. As an initial guess we choose 0.5\arcsec\ - corresponding to 10 (5) pixels on the PC (WF) chip. For some nearby galaxies, this turned out to be too small an area, and for such cases the procedure compares and rescales the fitting radius to correspond to a 30\,pc at the distance to the target. We use the distance modulus of the host galaxy obtained from the NED database (its median value entry). The choice of a minimum fitting radius of 0.5\arcsec\ also roughly corresponds to about three times the FWHM of the WFPC2 PSF. This radius is also used for the calibration of the filter zero points and transformation calibrations \citep{Holtzman95,Dolphin09}, which we later use in Section\,\ref{Sect: NSC photometry}. 

A second {\sc ishape} pass then adjusts the fitting radius to about three times the NSC major axis FWHM measured in the previous iteration. We allowed up to three iterations to avoid runaway and convergence effects. In practice, nearly all NSCs were fit with a fitting radius of $0.5\arcsec$, and only a few NSCs in the closest galaxies required a larger fitting radius.

Other free fitting parameters in {\sc ishape} are the cluster ellipticity and position angle. We allowed {\sc ishape} to compute the $r_{\rm eff}$ uncertainty from the correlated free parameters' uncertainties. While this makes the computation slower (by about a factor of three), it provides a more realistic estimate of the uncertainties in the derived $r_{\rm eff}$ values \citep{Larsen99}. For each galaxy image, our {\sc ishape} IRAF wrapper procedure fits subsequently each of the six 
analytical models, and selects the model with the lowest reduced $\chi^2$ residual. The $r_{\rm eff}$ values listed in Table\,\ref{Table:reff} are then calculated from the geometric mean value of the FWHM along the semi-minor ($\omega_x$) and semi-major ($\omega_y$) axis ($\sqrt{\omega_x\times\omega_y}$) and converted to $r_{\rm eff}$ using the conversion factors tabulated in the {\sc ishape} manual. 
Finally, the measured $r_{\rm eff}$ is calculated in parsecs using the distance modulus to the galaxy, which is given in Table\,\ref{Table:Galaxy sample}.

In some cases, the NSC light profile is fit equally well by different models. In these cases, we used a secondary metric to identify the best fitting model. Specifically, we used the residual (data - model) output images by {\sc ishape} (see Fig.\,\ref{fig:Residuals map}) to calculate the ratio between the standard deviation in the central $5\times5$\,pixels and that of the sky measured in an annulus with 3\,pixels width outside of the fitting radius. Essentially, this diagnostic measures the difference between the residuals and the local noise floor of the image\footnote{ This statistics quantifies the significance of the resi\-du\-als to the noise by comparing variances 
($\propto\!\sqrt{N_{\rm sky}}\times ( {\rm abs} (\sigma_{\rm residual}-\sigma_{\rm sky})/\sigma_{\rm sky})$) is similar to the \cite{Bartlett37} statistics testing for variances equality.}. 

The measured $r_{\rm eff}$ values for all NSCs and in all available filters are listed in Table~\ref{Table:reff}, while Table\,\ref{Table:ellpa} provides the ellipticities (1-b/a) and position angles (East of North) of the best-fitting NSC model. The latter was calculated using the {\sc ishape} position angle (measured clockwise from the detector y-axis) and the image header keyword ORIENTAT, which gives the position angle of the detector y-axis on the sky.

%%%%%%%%%%%%%%%%%%%%%%%%
\begin{figure}
\includegraphics[width=0.48\textwidth]{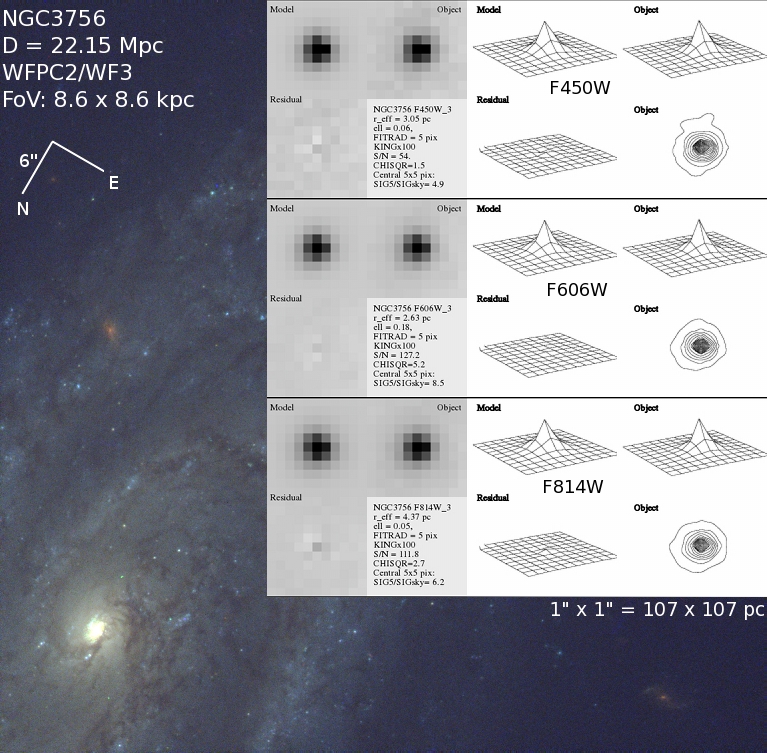}
\includegraphics[width=0.48\textwidth]{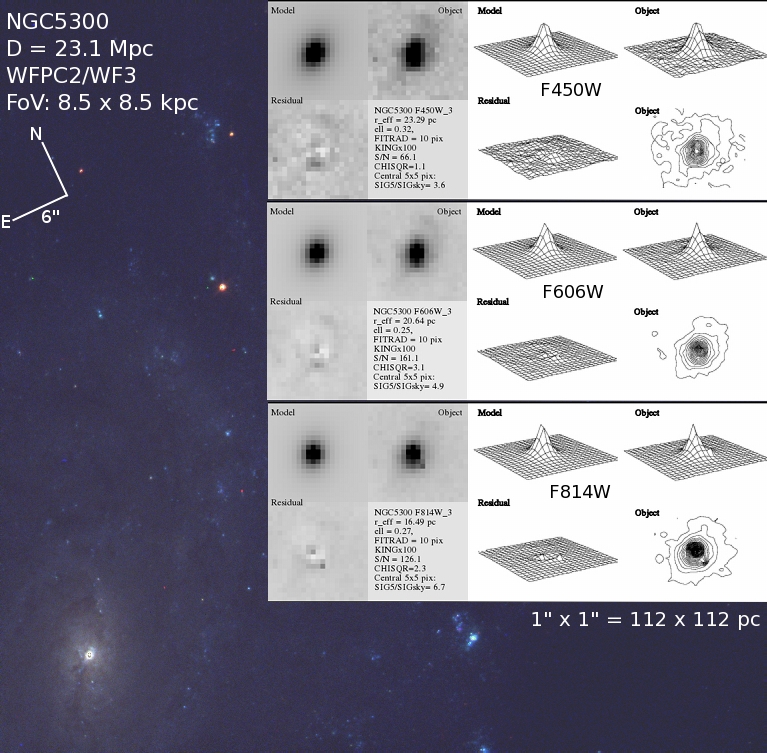}
\caption{Two examples for the results and data products: NGC\,3756 (top), a ``clean'' NSC, and NGC\,5300 (bottom), an NSC with faint circum-nuclear structures. The colour composite images are from the WFPC2/WF3 filters $F450W$, $F606W$, and $F814W$. The various inlet images contain gray-scale cut-outs and surface plots showing the results of the PSF-fitting for each filter. They all cover twice the {\sc ishape} fitting radius, and show the best-fit model, the object data, and the residual images (object-model) as labelled.}
\label{fig:examples}
\end{figure}
%%%%%%%%%%%%%%%%%%%%%%%%%%%%

Figure\,\ref{fig:examples} illustrates some examples for the results and data products of our PSF-fitting pipeline. The top panel shows the case of NGC\,3756, a fairly typical NSC with a ``clean'' circum-nuclear morphology, i.e. without significant residuals after the {\sc ishape} fitting. The bottom panel, in contrast, shows NGC\,5300, for which the residual images in all three filters reveal signs of various faint sources in the immediate vicinity of the NSC, most likely circum-nuclear star clusters and/or a small-scale star forming disc. This demonstrates the advantage of deriving the NSC properties from PSF-fitting, since other techniques (e.g. aperture photometry) would be influenced by such nearby contaminating structures, as discussed further in the next section.

Similar figures for the entire NSC catalogue are available in the online version of this article. Some example fits are shown in Fig.\,\ref{fig:Residuals map} in Appendix\,\ref{Appendix}.

\subsection{Obtaining NSC photometry}\label{Sect: NSC photometry}
%%%%%%%%%%%%%%%%%%%%%%%%%%%%%%%
Constraining the stellar population(s) of NSCs requires spectroscopy, or at least (multi) color information. While NSC spectroscopy with ground-based telescopes is, in principle, possible for nearby targets, it requires a significant time investment, even on 8m class telescopes \cite[e.g.][]{Walcher06,Seth10,Lyubenova13}. The sensitivity and resolution of the HST enables accurate and efficient multi-wavelength photometry for NSCs. The only systematic published survey of NSCs in  late-type spirals done with HST \citep{Boeker02} used only a single filter ($F814W$). This lack of precise multi-band photometry for NSCs for a large number of disk galaxies is one of the main motivations for our efforts to extract and analyse all such data available in the HST/WFPC2 archive. 

In general, aperture photometry of nuclear clusters in spiral galaxies is complicated due to contaminating light from various sources in their immediate vicinity (cf. Fig.\,\ref{fig:examples}). Under the assumption that the intrinsic light profile of the NSC is well represented by the best-fitting {\sc ishape} model, as described above, the flux contained in the model provides a more robust estimate of the NSC luminosity. Generally, we find that the best fit models leave relatively small residuals, typically $\lesssim5\%$ of the NSC flux (see Figure\,\ref{fig:examples} and Fig.\,\ref{fig:Residuals map}). 

We therefore derive the NSC magnitudes from the flux contained within the best fit {\sc ishape} model\footnote{This step is build in to our IRAF procedure, and takes place after the selection of the best-fit model. For the input list
of images and NSC coordinates, the procedure creates an ascii table containing the structural and photometric properties of the best-fit {\sc ishape} model (in instrumental and physical units), with one line per object listing the measurements for all available filters and detectors.}. 
For zeropoint and CTE correction for each exposure (taking into account the different detectors, gain settings, source positions and count levels, and epoch of observation), we use the values and prescriptions in \cite{Dolphin09}. We note that the majority of the NSCs, due to distance or smaller intrinsic size, were fit with an $0.5\arcsec$ fitting radius ($\approx3$x\,FWHM), which is identical to the aperture radius used by \cite{Dolphin09} to calibrate the transformation to instrumental magnitudes. Therefore, this guarantees minimal/negligible photometric calibration biases and uncertainties.

The resulting NSC magnitudes in all available filters are presented in Table\,\ref{Table: F300-814W_mag}. 
The listed photometric uncertainties are calculated using a local sky region outside the fitting radius and the calibration parameters uncertainties are added in quadrature.  We note here that these stochastic
uncertainties are often small compared to other, systematic, uncertainties which are discussed in Section\,\ref{Sect: Uncertainties}. In some cases, an NSC has been observed multiple times through the same filter by different observing programs and on different WFPC2 detector chips. In these cases, we give priority to the highest $S/N$ observation, and indicate the WFPC2 detector with a subscript to the listed magnitudes.

To facilitate comparison with ground-based studies, we also transformed the WFPC2 magnitudes in Table\,\ref{Table: F300-814W_mag} to the Johnson-Cousins magnitude system, which we give in Table\,\ref{Table: UBVI_mag}. For this transformation, we use the \cite{Dolphin09} coefficients and the measured NSC colour, if available. When there is no colour information (i.e. for a single WFPC2 filter in the HST archive or other were saturated), we assume the colour of a \cite{BC03} SSP model for an age of 5\,Gyr and solar metallicity. While these are reasonable assumptions for NSCs \citep{Walcher06}, the Johnson-Cousins magnitudes for NSCs with no measured colour information should be used with caution. 

Because \citeauthor{Dolphin09} does not provide transformation coefficients to $U$-band, we adopt those from \cite{Holtzman95} for both $F336W$ and $F300W$. For the bluest HST filters, \citeauthor{Dolphin09} finds only small differences to the \cite{Holtzman95} calibration. The $U$-band magnitudes in Table\,\ref{Table: UBVI_mag} should therefore be fairly accurate. Nevertheless, the $U-$band magnitudes should also be used with care for any comparative analysis, and the native WFPC2 magnitudes should be used instead. 

For each galaxy, we retrieved (from NED) the
foreground Galactic extinction based on the \cite{Schlafly11} 
recalibration of the \cite{Schlegel98} extinction map, and calculated filter-specific values assuming the \cite{Fitzpatrick99} reddening law with $R_V=3.1$. 

\subsection{Uncertainties, limitations, and quality checks}
\label{Sect: Uncertainties}
%%%%%%%%%%%%%%%%%%%%%%%%%%%%%%%

\begin{figure*}
\epsfig{file=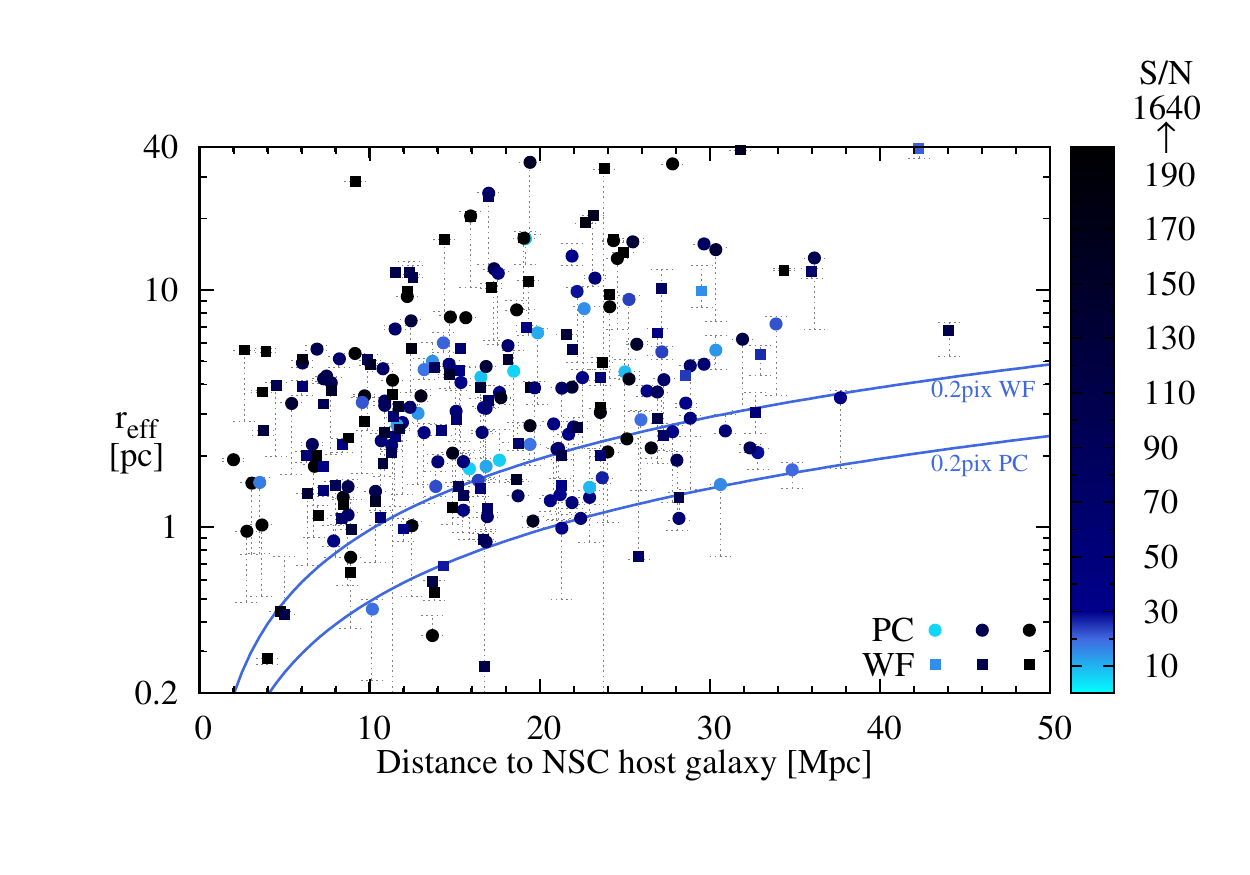, width=0.95\textwidth, bb=20 30 360 230}
\caption{NSC effective radii as a function of distance to the host galaxy. Circles indicate NSCs observed with the PC detector, while squares are for those located on the WF detectors. Each symbol is colour-coded according to the $S/N$ of the NSC, as indicated by the vertical colour bar. The solid curves show the limits to which an NSC can be recognized as an extended source, provided it has $S/N\!\geq\!30$ (see \S\ref{Sect: Uncertainties}).
The majority of NSCs fall above the curve for their respective WFPC2 detector, i.e. they are resolved, and their $r_{\rm eff}$ measurements are reliable.}
\label{fig: reff_dist}
\end{figure*}

As discussed in Sect.\,\ref{Sect: Measuring NSC sizes}, a reliable measurement of $r_{\rm eff}$ relies on a well characterized instrument PSF, and on data with a sufficiently high signal-to-noise ratio \cite[$S/N>30$ for the case 
of compact star clusters,][]{Larsen99}. To better gauge the reliability of the measured 
$r_{\rm eff}$ for our sample of NSCs, we show in Figure\,\ref{fig: reff_dist} 
the measured NSCs' $r_{\rm eff}$ as a function of distance to the host galaxy. The solid curves indicate the smallest cluster size, i.e. the smallest measurable difference from the instrument PSF, that can be reliably measured for objects with $S/N>30$ for given WFPC2 detector. For a well-sampled PSF, this limit corresponds to 10\% of the PSF width, or 0.2 pixels on the PC chip, as indicated by the lower curve. Because of the coarser sampling of the WF chips, this accuracy is difficult to achieve for NSCs observed on these detectors. For these to be considered as ``resolved'', we thus adopt a conservative limit of 20\% of the PSF width (0.2 pixels on the WF chip), as indicated by the upper curve.
The symbols for the NSC measurements have been color-coded to highlight those observations that have marginal $S/N$ (light blue symbols), and therefore may not reach these resolution limits. 

Figure\,\ref{fig: reff_dist} demonstrates that nearly all NSCs in our sample fall above their respective curve, i.e. they are well-resolved and the {\sc ishape} fits provides a robust measurement of their effective radii. Those NSCs which fall below the resolution limit of their respective detector and/or have $S/N<30$,
are regarded as upper limits, marked accordingly in Table 3, and are not considered for the following analysis. The remaining 202 NSCs (89\% of the sample) are well-resolved, and their effective radii measurements should therefore be reliable.

\begin{figure}
\epsfig{file=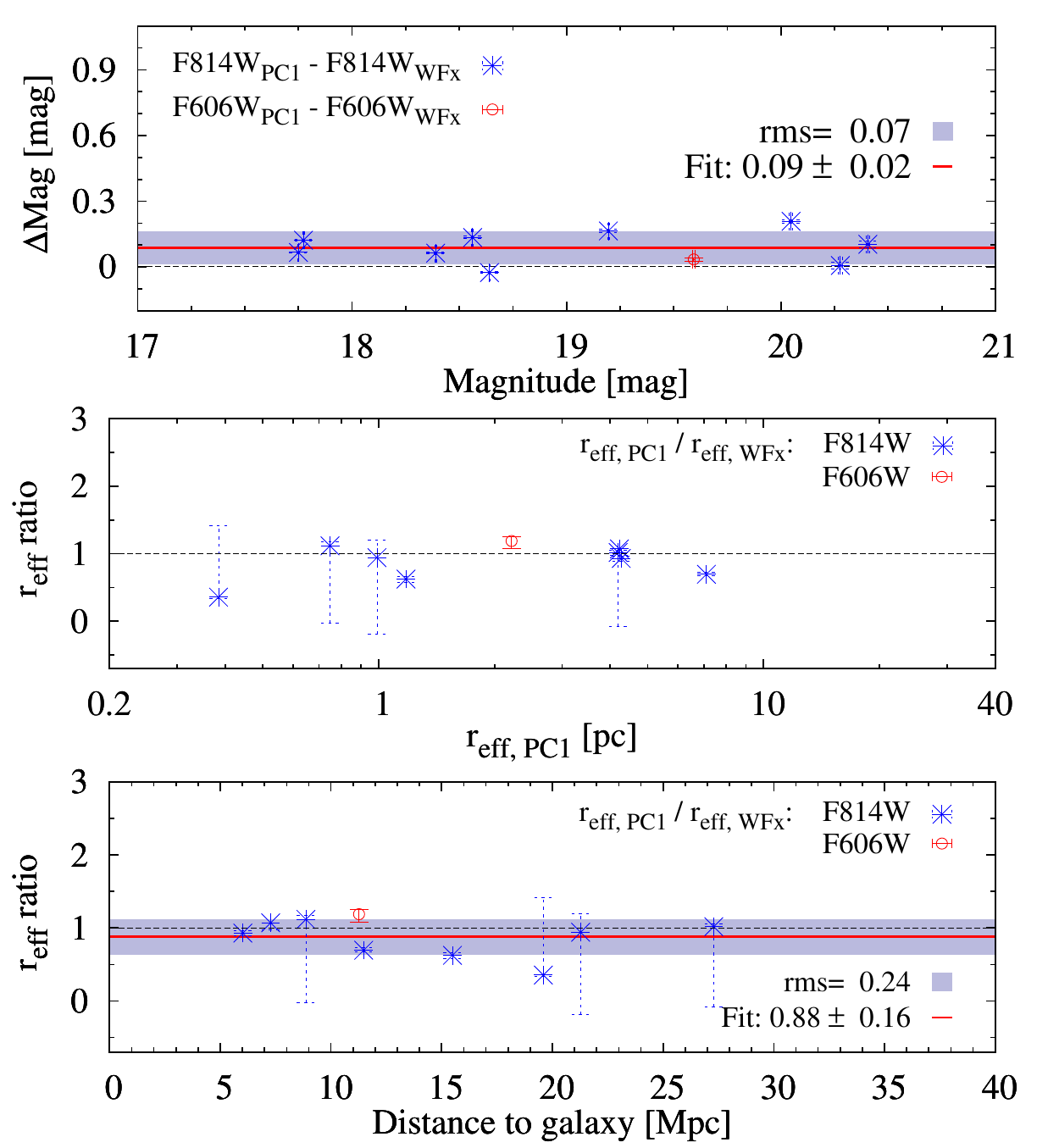, width=0.5\textwidth, bb=20 5 360 380}
\caption{Comparison of NSC measurements obtained with the same filter, but on different WFPC2 detectors. The three panels show the luminosity difference (top), and the $r_{\rm eff}$ ratios as a function of cluster size (middle) and distance to the NSC host galaxy (bottom). The solid lines in the top and bottom panels denote the error-weighted least square fits to the data, while the gray bands indicate the standard deviation.
}\label{fig: delta_mag_reff_ratio}
\end{figure}

In order to demonstrate that our $r_{\rm eff}$ measurements are indeed robust against differences in the spatial sampling of the WFPC2 detectors, we compare in Figure\,\ref{fig: delta_mag_reff_ratio} 
the results obtained for NSCs that have been observed on both the PC and WF chips, fall above the resolution limit, and have $S/N\geq30$. This selection strategy is adopted for all plots, unless indicated otherwise.
Unfortunately, there are only eight NSCs that were observed in 
the same filter, but on two different WFPC2 detectors. For these, we plot the difference in the derived magnitudes (top panel), as well as the ratio of the two measured $r_{\rm eff}$ values against both $r_{\rm eff}$ (as measured on the PC chip, middle panel) and the distance to the host galaxy (bottom panel). The $r_{\rm eff}$ ratios scatter around unity (with a mean $r_{\rm eff}$ of 0.88 and an rms of 0.24), independently of cluster size or distance. Although low-number statistics prevents a more quantitative discussion, it is worth noting that the somewhat smaller $r_{\rm eff}$ values on the PC detector indicated by the fit are not unexpected given the lower spatial resolution of the WFx measurements.

The top panel of Figure\,\ref{fig: delta_mag_reff_ratio} shows that the NSC photometry, on the other hand, appears to depend somewhat on the spatial resolution of the data, in the sense that the NSC is fainter by about 0.1 mag when observed with the PC chip. In general, such an effect is quite plausible in the regime of marginally resolved sources because the central pixel on the WF chips may already contain flux from circum-nuclear structure. However, the small number of objects with observation on both detectors (and the lack of similar information for other filters) prevents us from a quantitatively reliable correction across the NSC sample. Instead, we %conservatively 
adopt the rms variation between different observations of the same objects through the same filter (0.07 mag) as the minimum uncertainty for all NSC photometric measurements. This uncertainty is used to derive the typical errors for the NSC colour shown in Figures\,\ref{fig:WFPC2_CCD} and \ref{fig:NSCs CMD}.

\begin{figure}
\epsfig{file=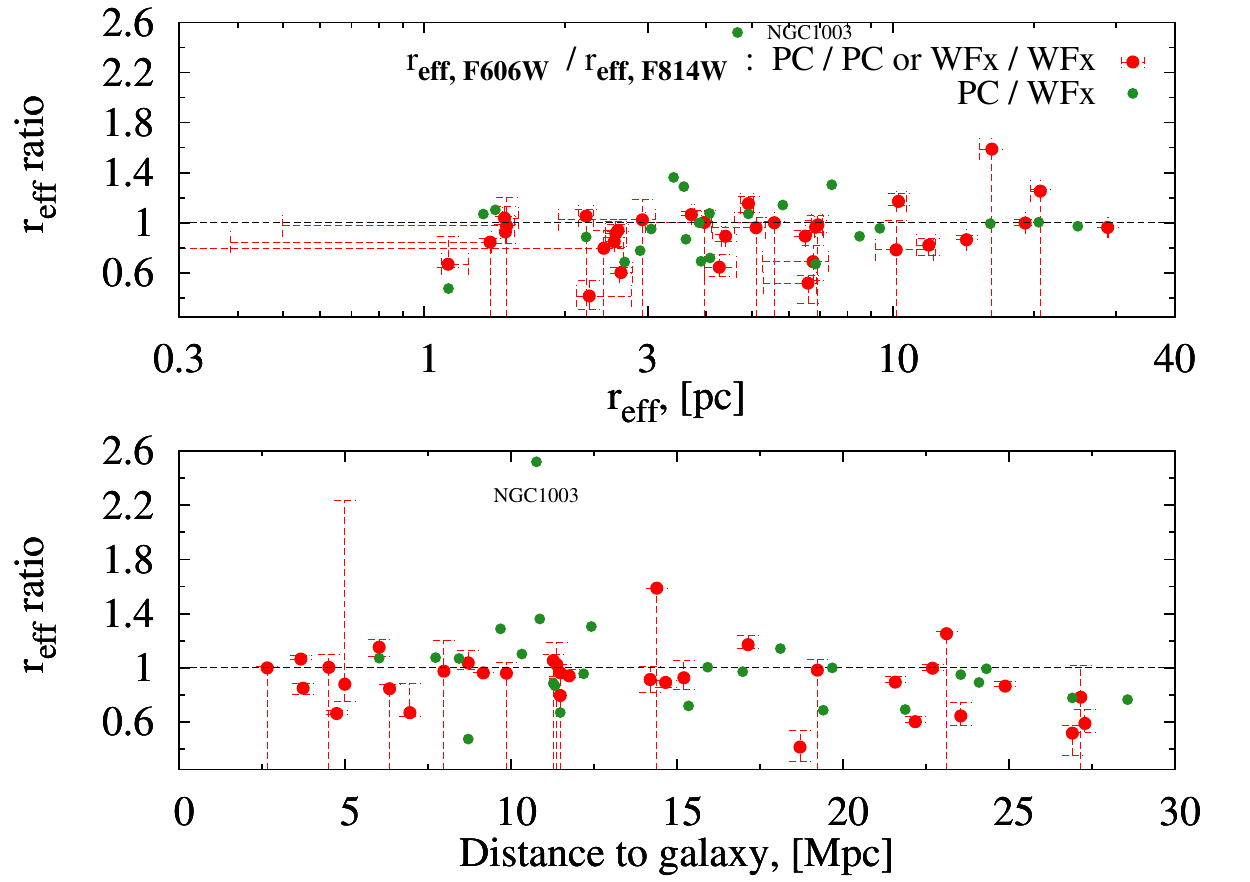, width=0.58\textwidth, bb=10 5 410 250}
\caption{Ratio between $r_{\rm eff}$ values measured in two different filters on the same (red symbols) or on different WFPC2 detectors (green symbols), plotted against NSC size (top) and distance to the host galaxy (bottom).
}
\label{fig:reff_blue_to_reff_red}
\end{figure}

In order to check for any dependence of $r_{\rm eff}$ on wavelength, we show in Figure\,\ref{fig:reff_blue_to_reff_red} the ratios of $r_{\rm eff}$ for all NSCs observed in both $F606W$ and $F814W$. While there is no obvious trend with NSC size (top panel) or galaxy distance (bottom panel), the average ratio seems to be smaller than one, i.e. NSCs appear to be slightly larger in $F814W$ than in $F606W$. While we will discuss this issue in more detail in \S,\ref{Sect: Size Distribution}, we point out here that the presence of star formation in the immediate vicinity of the NSC and the associated $H_\alpha$ emission will cause the opposite effect, i.e. it will {\it increase} the apparent NSC size measured in the $F606W$ filter, since for nearby galaxies, the $F606W$ passband includes the $H_\alpha$ line. 

\begin{figure}
\includegraphics[width=.5\textwidth]{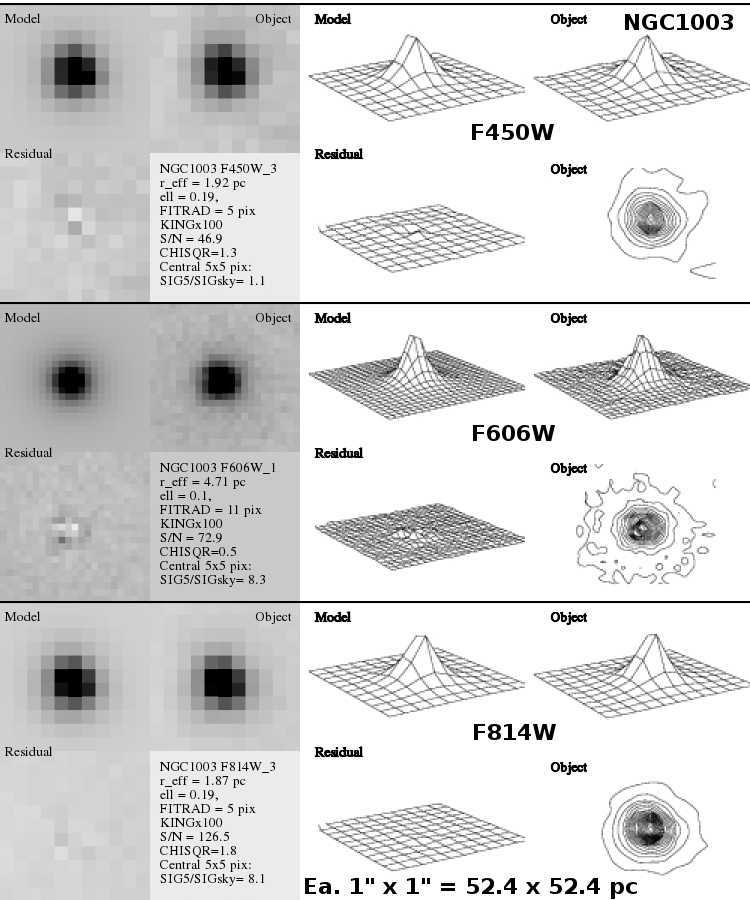}
\caption{NGC\,1003 profile fitting maps in all three WFPC2 filters, showing the excess flux and structure visible in $F606W$. Each panel shows image, surface and contour plot of the best-fit model (left), the object (right) and the residual (data - model, bottom left). 
\label{fig: NGC1003_rsd_maps}
}
\end{figure}

As a case in point, the only outlier in Figure\,\ref{fig:reff_blue_to_reff_red} is NGC\,1003, a nearby galaxy with a high $S/N$ NSC. Its $F606W$ image (obtained with the higher resolution PC1 detector) shows a faint extended structure evident in the {\sc ishape} fit residuals shown in Figure\,\ref{fig: NGC1003_rsd_maps}. It is likely that this structure is caused by (circum-)nuclear star formation, and that it is responsible for the much larger $r_{\rm eff}$ measured in the  $F606W$ image. Indeed, \cite{Moustakas06} have observed an $H_\alpha$ flux of $F_{H_\alpha}=1.45$ mW/m$^2$ from the central $2\arcsec.5\times2\arcsec.5$ ($132\times132$\,pc) of NGC\,1003. 

As described in \S\,\ref{Sect: Measuring NSC sizes}, we assume an intermediate-type SED when generating the PSF model with {\sc TinyTim}. In principle, the choice of the object spectrum impacts the PSF model for a given filter passband. In order to quantify the maximum error that may be caused by the SED choice, we derived the effective radius of two NSCs with different colours (one blue, one red) using {\sc tinytim} PSF models generated with both a hot (spectral type F) and a cold (spectral type M) SED. We find that the $r_{\rm eff}$ values remain within the measurement uncertainty ($<\!10$\% scatter in $r_{\rm eff}$) regardless of the 
choice of PSF color. This is in agreement with previous studies \cite[e.g.][see also Sect.\,\ref{Sect:Comparison to Previous Studies}]{Boeker04}. On average, we find that colours can be affected by $<\!0.1$\,mag (in either direction) if the shape of the PSF spectrum does not match that of the NSC. We therefore consider the assumption of an intermediate-type input spectrum to be an acceptable compromise.

\begin{figure}
\epsfig{file=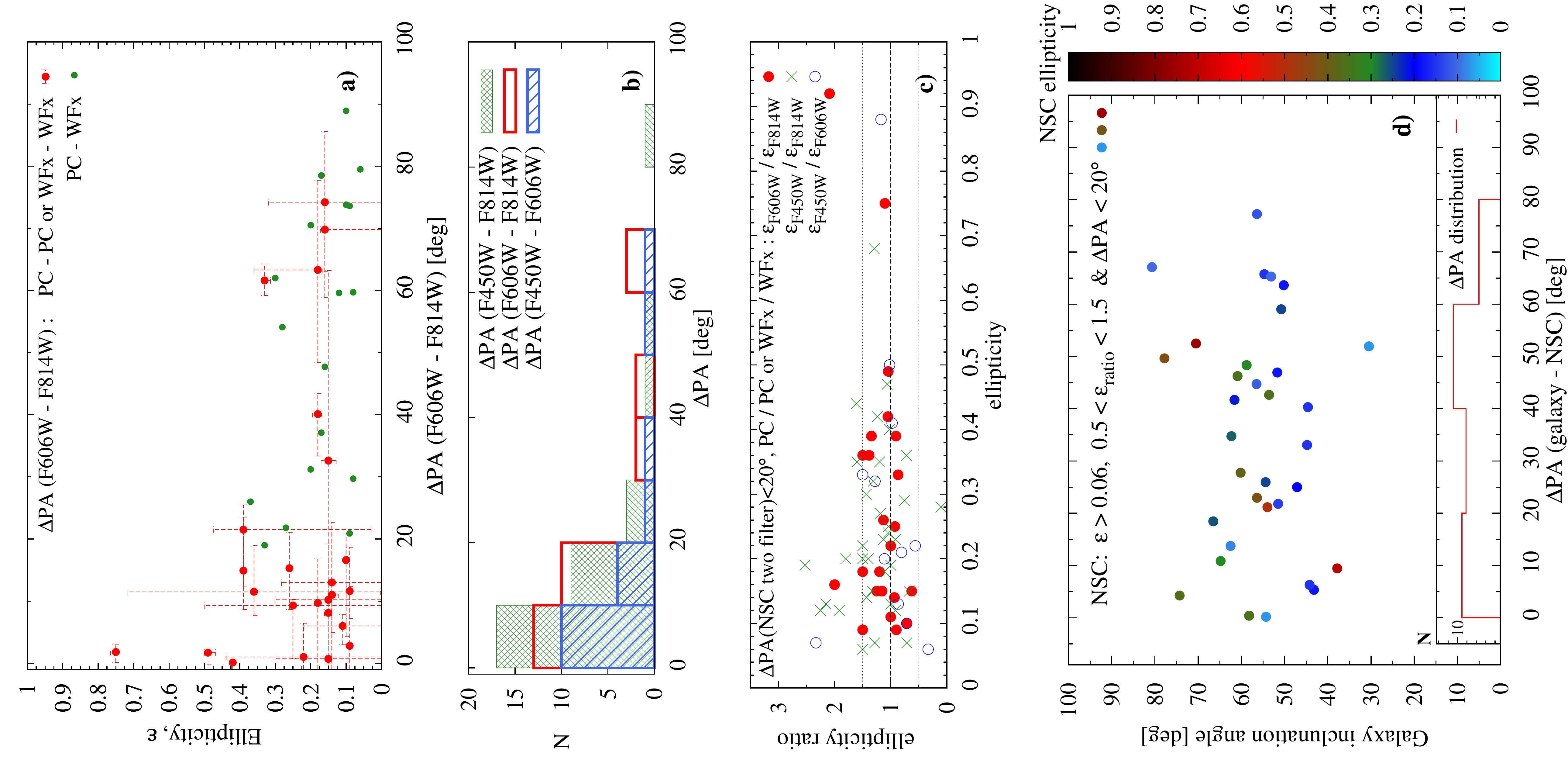, angle=-90, width=1\textwidth, bb=10 10 1470 1470}
\caption{a) NSC ellipticity ($\epsilon$, in $F606W$) against the difference of the NSC PA values measured in two different fil\-ters, $\Delta$PA($F606W\!-\!F814W$), b) Histogram of $\Delta$PA %, only NSCs in the first bin are considered reliable 
for the indicated filter pairs, c) ratio of $\epsilon$ values measured in the respective filter pairs. Only objects in the first two bins of panel b) are plotted. We consider only NSCs with eliipticity ratios between 0.5 and 1.5 (i.e. between the two dotted lines) as reliable. d) difference between the PAs of NSC (the average of the respective filter pairs) %values measured in $F606W$ and $F814W$) 
and the host galaxy disk, plotted against galaxy inclination, and color-coded by NSC ellipticity. Only objects with trustworthy measurements of PA and $\epsilon$ are plotted. The histogram in panel d) shows the respective $\Delta$PA distribution.
\label{fig:NSC_Gal_PA}}
\end{figure}

Lastly, in Figure\,\ref{fig:NSC_Gal_PA} we examine the robustness of the derived shape parameters of NSCs (i.e. position angle PA and ellipticity $\epsilon$) by comparing their measurements in different filters. The top panel of Fig.\,\ref{fig:NSC_Gal_PA} shows the value of $\epsilon$ measured in $F606W$ (for NSCs with $\epsilon > 0.06$) against the difference between the PA values measured in the $F606W$ and $F814W$ filters. There is a rather large scatter in the PA difference, especially for ellipticities below 0.2, indicating that PA measurements for mostly round NSCs are expectedly unreliable. We therefore focus our analysis on those NSCs for which the PA difference is smaller than $20^{\circ}$, i.e. the first two bins of the histogram plotted in the second panel. For these clusters, the ellipticity ratio between the two filters is plotted in the third panel. Since there are still many measurements that do not agree well between filters, we apply another selection by focussing on those that agree to within 50\% between different exposures. 

The bottom panel of Figure\,\ref{fig:NSC_Gal_PA} then compares the PA of the NSC with the PA of the host galaxy disk. Here, we show only NSCs meeting the above criteria for a credible measurement of both PA and $\epsilon$. This is an important test because a number of studies \cite[e.g.][]{Seth06,Hartmann11} have suggested that a general alignment between NSC and host galaxy disk can be expected if the NSC primarily grows via gas accretion, rather than the infall of star clusters. However, it is evident from the bottom panel of Figure\,\ref{fig:NSC_Gal_PA} that there is no general alignment between NSC and host galaxy disk. 
%We caution, however, that 
However, this observation does not imply that gas accretion is ruled out. There are 
%individual objects 
known galaxies for which the NSC is well aligned with the disk, mostly in edge-on spirals \citep{Seth06}, NGC\,4244 \citep{Seth08b}, in the Milky Way \citep{Schoedel14} or NGC\,4449 (see \S\,\ref{Sect: multiple NSCs}). Our result merely suggest that this is not universally the case. It is possible that the initial information about orientation alignment may be erased for a more dynamically evolved NSCs due to internal dynamical evolution, which operates on a cluster relaxation time scale. Projection effects and NSC triaxiality may be another factor at play.

\subsection{Comparison to previous studies}\label{Sect:Comparison to Previous Studies}
%%%%%%%%%%%%%%%%%%%%%%%%%%%%%%%

\begin{figure}
\epsfig{file=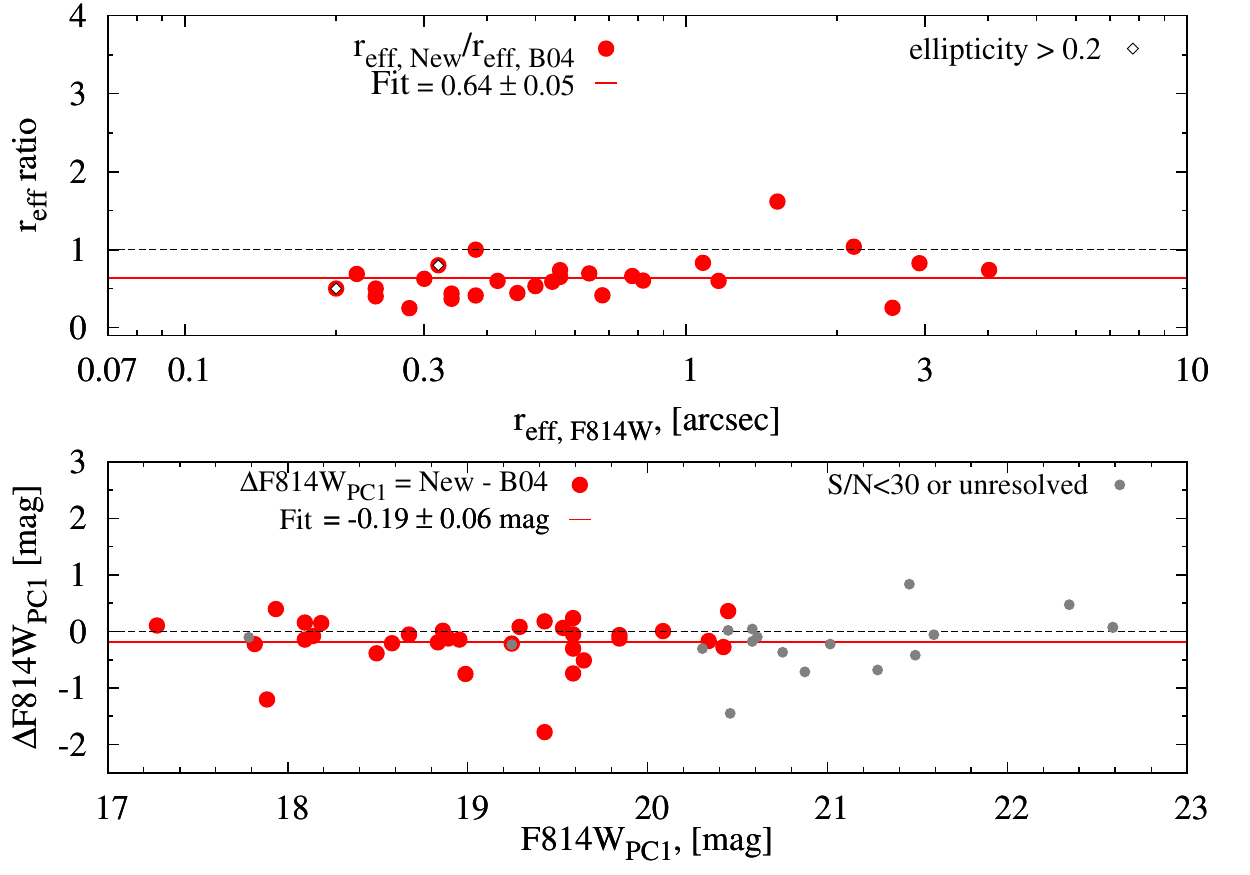, width=0.5\textwidth, bb=5 10 360 260}
\caption{Comparison of our NSC size and magnitude measurements to those of \protect\cite{Boeker04}. The top panel shows the ratio between the $r_{\rm eff}$ measured in the two studies, while the bottom panel shows the difference in the apparent F814W magnitude. The grey dots in the bottom panel denote NSCs that only have upper limits for their effective radii as described in \S\,\ref{Sect: Uncertainties}. Open diamonds indicate NSCs with ellipticity $\epsilon>0.2$. The red line in each panel shows the results from the least squares fit (to the red circles only). 
}\label{fig: d_reff with B04}
\end{figure}

In order to further gauge the reliability of our measurements, we compare in Figure\,\ref{fig: d_reff with B04} the NSC sizes and magnitudes measured in this work with those of 39 NSCs in \cite{Boeker04} that were derived using the same WFPC2 data. The top panel shows the ratio between the two NSC radius measurements, while the bottom panel shows the difference in their measured magnitudes. All measurements are compared in arcseconds and apparent magnitudes, respectively, in order to avoid differences due to adopted distances and Galactic reddening values between the two studies. 

The comparison reveals that, on average, the newly measured radii are smaller by about 35\% than those in \cite{Boeker04} (Fig.\,\ref{fig: d_reff with B04}, bottom panel). Part of this difference can be explained by the improved PSF fitting technique used in this work, which allows for non-spherical NSC shapes. For example, an NSC ellipticity of $\epsilon\simeq0.2$ would yield an $r_{\rm eff}$ value that is smaller by 10\% when using the geometric mean instead of the semi-major axis value, as adopted by \cite{Boeker04}. On the other hand, only very few NSCs in our sample have ellipticities above $\epsilon>0.2$ 
(see Figure\,\ref{fig: d_reff with B04}, top panel).

The remainder of the systematic difference is likely explained by the fact that the present work adapts the fitting radius to the object FWHM. This additional step makes it less likely that the fit is performed on a too large area which may include circum-nuclear structures, and therefore avoids a bias towards larger NSC sizes. We verified this on a few images by enforcing a stepwise increase of the fitting radius, i.e. from 5\,pixels to 11\,pixels, and found that the measured $r_{\rm eff}$ can easily increase by 10\% if the fitting radius is too large. Other differences to the work of \cite{Boeker04} that may contribute to the differences include the use of improved software versions of both the {\sc TinyTim} code for generating the WFPC2 PSF models and the {\sc ishape} code itself. 
We conclude that the systematically smaller NSC sizes reported here can be explained by the various improvements in the fitting method, and that the new $r_{\rm eff}$ measurements are more accurate.

As for the photometric comparison (Fig.\,\ref{fig: d_reff with B04}, bottom panel), the agreement with the earlier work is generally better. Here, we also include those NSCs that are only marginally resolved and/or have a low S/N ratio (gray dots). On average, the newly measured apparent magnitudes are somewhat brighter (by about 0.2\,mag) than those measured by \cite{Boeker04}. This difference can easily be explained by recent improvements to the WFPC2 calibration, especially the correction for charge-transfer-efficiency (CTE), which are included in the \cite{Dolphin09} zeropoints used for this work, but not in the \cite{Holtzman95} calibrations used by \cite{Boeker04}. There are a few sources with rather large discrepancies of one magnitude or more; these are always in the direction of the new measurements being brighter. We attribute these to the improved fitting technique which avoids fitting errors due to emission from circum-nuclear structures. 

%%%%%%%%%%%%%%%%%%%%%%%%%%%%%%%%%%
\section{Results and Discussion}\label{Sect:Results and Discussion}
%%%%%%%%%%%%%%%%%%%%%%%%%%%%%%%%%%

In this section, we present the distributions of effective radii, luminosities, and colours for the full NSC catalogue. We discuss the properties of NSCs 
in comparison to those of other compact stellar systems (GCs, UCDs, and NSCs in early-type galaxies), and briefly address some proposed scenarios for their formation. In particular, we highlight in \S\,\ref{Sect: multiple NSCs} an example for each of the two most likely scenarios for the growth of NSCs, namely cluster merging and {\emph in situ} star formation.

\subsection{Size Distributions}\label{Sect: Size Distribution}
%%%%%%%%%%%%%%%%%%%%%%%%%%%%%%%%%%

\begin{figure}
\epsfig{file=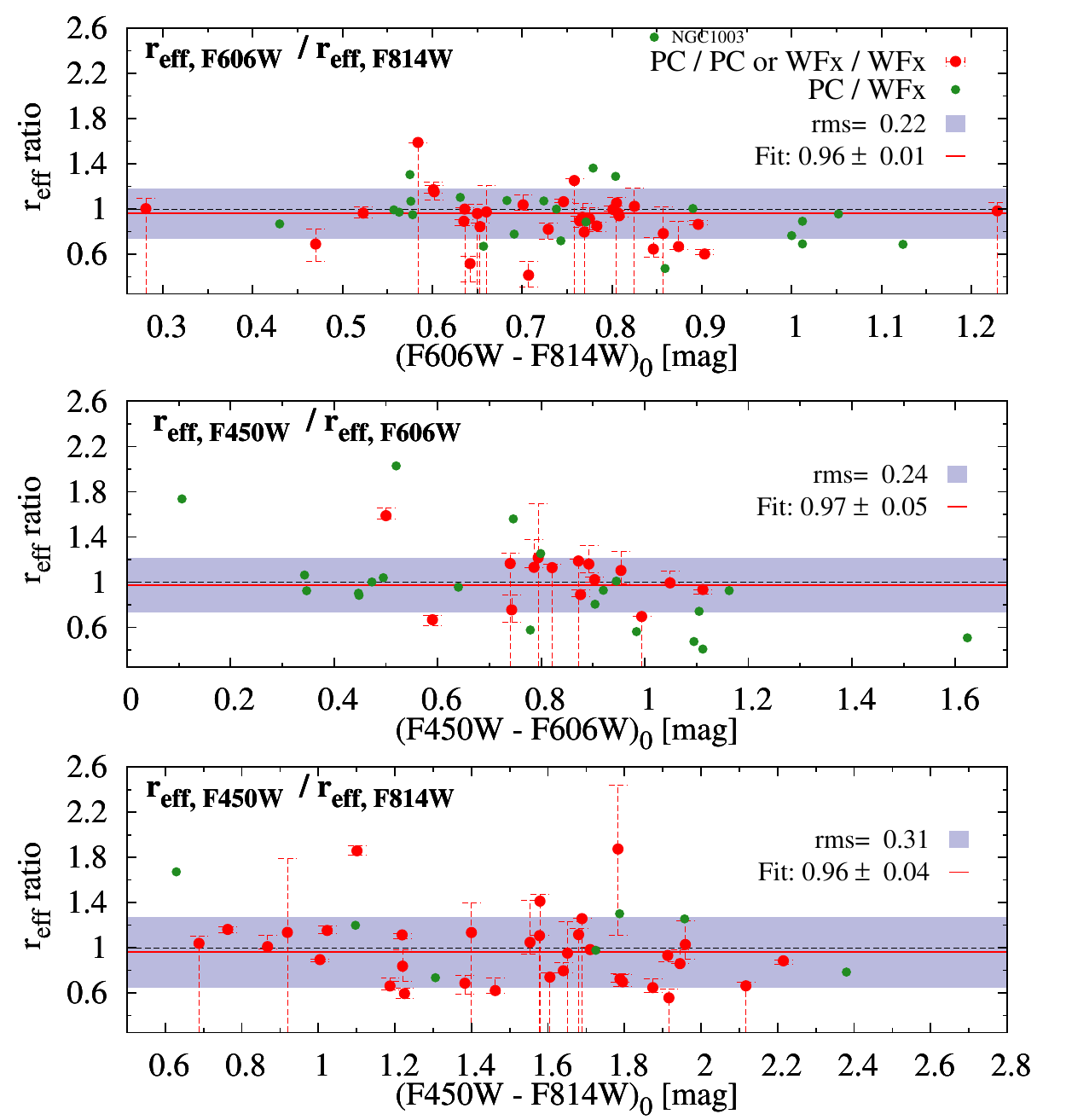, width=0.58\textwidth, bb=10 5 410 370}
\caption{Ratio between $r_{\rm eff}$ values measured in two different filters on the same (red symbols) or on different WFPC2 detectors (green symbols), plotted against the three most common NSC colours. Only $r_{\rm eff}$ ratios for well-resolved NSCs with $S/N\geq30$ are shown. The error-weighted least-squares fit to the size ratios are shown with solid (red) line and the rms scatter is indicated by the horizontal band. The fit statistics are shown in each panel.
}
\label{fig:reff_ratios}
\end{figure}

In \S\,\ref{Sect: Uncertainties}, we noticed that NSCs appear slightly smaller when observed in bluer filters. This is a potentially important result because it could indicate radial stellar populations variations within NSCs, and thus merits a more detailed investigation. To quantify this effect further, we examine in Figure\,\ref{fig:reff_ratios} how the $r_{\rm eff}$ measurements in the three most common filters compare to each other. While the size ratios in all three filter combinations scatter around unity (dashed line), without any systematic trend with colour, the average size ratio is slightly, but nevertheless significantly, smaller than one, as indicated by the error-weighted least-squares fits in the three panels of Figure\,\ref{fig:reff_ratios}.

To test the statistical significance of this result, we use the Wilcoxon test within the $R$ software package\footnote{{\sc R} is a free software environment for statistical computing. The R-project is an official part of the Free Software Foundation's GNU project. http://www.r-project.org/},
which is superior to other paired difference rank tests because it does not assume that the data are normally distributed. 
The test uses the data median to test whether it is consistent with the null hypothesis, which in our case is that the median is identical to the derived fit value.
For all three filter combinations, we find that the null hypothesis is true with a high probability, i.e. the $p$-values are greater than 0.9 for all three filter combinations. In contrast, $p\leq0.2$ for the hypothesis that the median size ratio is equal to one. 

\begin{figure}
\epsfig{file=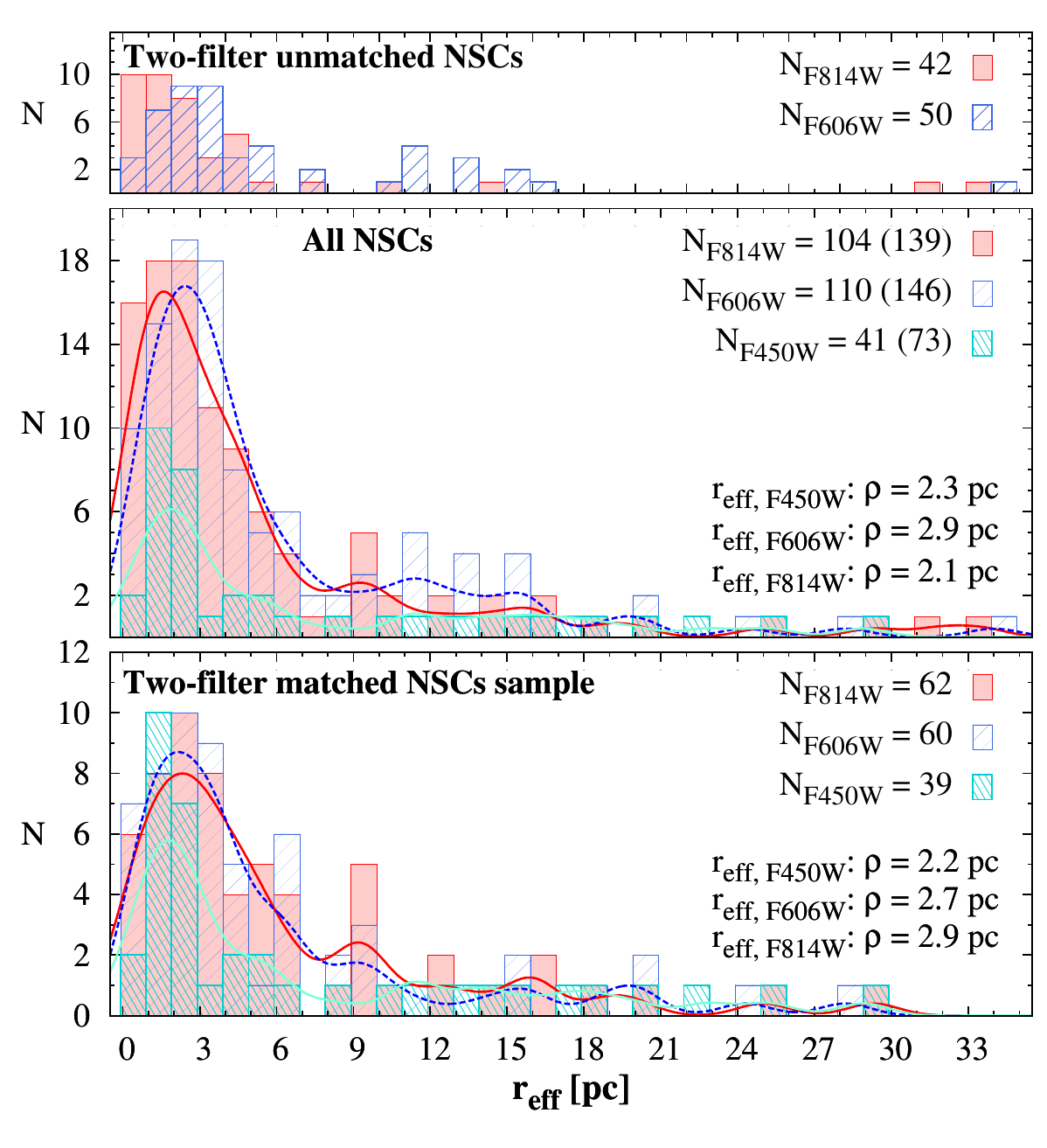, width=1\textwidth, bb=20 15 700 370}
\caption{
Histograms of NSC effective radius for the three most commonly observed WFPC2 filters. The middle panel shows the $r_{\rm eff}$ distributions for all resolved NSCs in our sample with high S/N. The bottom panel only includes those NSCs that have been observed in both $F606W$ and $F814W$, so that the sample is identical to the one in Fig.\,\ref{fig:reff_ratios}. The top panel only plots those NSC with data in either $F606W$ or $F814W$ (but not both). The respective sample sizes are shown in the legend. The solid (dashed) curve in the lower two panels shows the $F606W$ ($F814W$) probability density estimates, with their peak values listed in the respective panel.
\label{fig: WFPC2_reff_hist_BVI} }
\end{figure}
Another illustration of this results is shown in Figure\,\ref{fig: WFPC2_reff_hist_BVI} which shows the $r_{\rm eff}$ distribution measured in the three most common WFPC2 filters. The first point to make is that Figure\,\ref{fig: WFPC2_reff_hist_BVI} confirms results from previous studies \citep{Boeker04,Cote06} that the $r_{\rm eff}$ distribution peaks around 3\,pc, and has a long tail towards larger radii. 

In the bottom panel, we plot only those NSCs that were observed in the {\emph same} matched filter pairs as
those in Figure\,\ref{fig:reff_ratios}.
To check for systematic differences in NSC size with filter passband, we perform a non-parametric probability density estimation on the $r_{\rm eff}$ data (within {\sc R}) with a kernel window of 1\,pc (twice the typical $r_{\rm eff}$ uncertainty). The same bin width is adopted for the histogram representation. It confirms that on average, the $r_{\rm eff}$ distribution peaks at a slightly smaller value in $F606W$ than in $F814W$. The difference is about 7\%, consistent with the results of Figure\,\ref{fig:reff_ratios}. The trend is confirmed also when comparing to the $F450W$ distribution, which shows an even smaller average NSC size, albeit with less statistical significance due to the smaller sample size.

Note, however, that for the full NSC catalogue (middle panel), 
the histogram for $F606W$ appears to peak at a slightly {\it larger} radius than those for $F450W$ and $F814W$, which seems to contradict the above result. This apparent contradiction is explained by the fact that the full sample plotted in the middle panel includes many NSCs that are observed only in a single filter. As the top panel of Figure\,\ref{fig: WFPC2_reff_hist_BVI} illustrates, the NSCs that are observed only in $F606W$ are on average larger than those observed only in $F814$.

The reasons for this prominent selection effect are not immediately obvious. As discussed in \S\ref{Sect: Uncertainties}, the presence of $\rm H_\alpha$ emission in the immediate vicinity of the NSC can affect the {\sc ishape} fits in the F606W filter such that the measured $r_{\rm eff}$ value is larger in this filter. While we cannot easily deduce the selection criteria for archival data, it is possible that galaxies observed only in $F606W$ are more likely to show $\rm H_\alpha$ emission than those observed only in $F814W$, thus explaining their larger average size.

There are a couple of plausible explanations why NSCs may appear smaller in bluer filters. The first explanation that warrants consideration is 
%the higher spatial resolution at shorter wavelengths. Even though we exclude unresolved NSCs and the upper limits for their $r_{\rm eff}$ from Figures\,\ref{fig:reff_ratios} and \ref{fig: WFPC2_reff_hist_BVI}, it is possible that some NSCs are only marginally resolved in $F814W$, thus causing their sizes to be slighly overestimated in that filter. The second possibility is 
that the smaller NSC sizes in bluer filters reflect real differences in the shape of the NSC. One 
possibility is the presence of a weak AGN, which would add a blue, unresolved source to the NSC profile, thus making it appear more compact at shorter wavelengths. Another plausible explanation are radial variations in the dominant stellar population within a given NSC, in the sense that the young (blue) population is more concentrated than the old population. This scenario is plausible if NSCs grow from infalling gas %clumps via 
and experience recurrent {\emph in situ} star formation. A conclusive analysis as to what effect is responsible requires a more detailed case-by-case analysis of some well-resolved NSCs, which is beyond the scope of this paper.

\begin{figure}
\epsfig{file=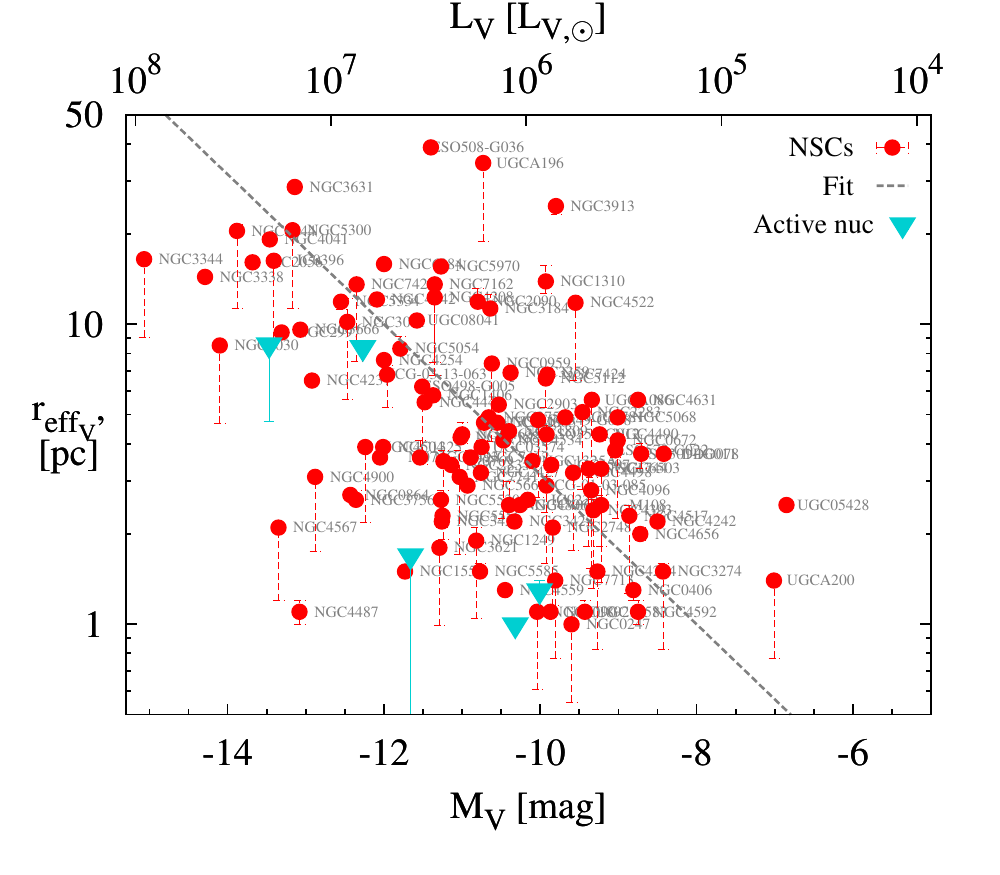, width=0.5\textwidth, bb=10 15 280 252}
\caption{Size-luminosity relation for the NSC catalogue presented in this study. Only well-resolved sources with high S/N are shown, as discussed in \S\,\ref{Sect: Uncertainties}. Dashed line shows a fit to the size luminosity distribution (see eq.\,\ref{eq:size-mass}). With triangles we show a control subsample of nuclei in active galaxies (see \S\,\ref{Sect: AGNs}).}
\label{fig: WFPC2_reff_Mv_BVI_all}
\end{figure}
%
%%%%%%%%%%%%%%%%%%%%%%%%%
\subsection{Size-luminosity Relation}
%%%%%%%%%%%%%%%%%%%%%%%%%
%
In Figure\,\ref{fig: WFPC2_reff_Mv_BVI_all}, we show the size-luminosity distribution of the NSCs in our sample.
The absolute $V-$band magnitude, $M_V$, is derived from the $F606W$ (or $F555W$) magnitudes of the best-fit {\sc ishape} model as explained in \S\,\ref{Sect: NSC photometry}. The inverted triangles mark a small control sample of NSCs harbouring a weak AGN (see more details in \S\,\ref{Sect: AGNs}). For these, the measured $r_{\rm eff}$ should be considered as upper limits, most likely because unresolved emission from the AGN makes the NSC profile indistinguishable from a point source.

It is evident from Figure 9 that NSCs follow a size-luminosity relation. The least squares fit to the data yields the following relations:
\begin{equation}
\log r_{\rm eff} = -2.0\pm0.2 - 0.25\pm0.02\ M_V\label{eq:size-mass}
\end{equation}
or
\begin{equation}
r_{\rm eff} = 10^{-3.21\pm0.3}\times L_V^{0.625\pm0.11} .
\end{equation}
The effective radii of NSCs scale with luminosity as a power law with an exponent of 0.625, which is very similar to that derived by \cite{Evstigneeva08} for a sample of UCDs as well as to the one measured for a sample of dE nuclei by \cite{Cote06}. At least in this context, UCDs and the NSCs in spirals and dE galaxies appear to share similar evolutionary histories, a topic that is discussed further in \S\,\ref{Sect: NSCs_GC_UCDs}.

The observed scatter in Figure\,\ref{fig: WFPC2_reff_Mv_BVI_all} is much larger than what can be explained by uncertainties in the photometric transformations ($\leq\pm0.1$\,mag) and/or distance modulus ($\leq\pm0.2$\,mag) which can only cause variations in $r_{\rm eff}$ of $\leq\pm20$\% and $M_V$ of $\leq\pm0.3$\,mag. 
This demonstrates that the large spread in the size-luminosity relation of NSCs is caused by variations in their M/L ratio, which may be indicative of differences in their evolutionary stage. For example, passive ageing alone can already change the $V$-band magnitude of an NSC by about a 1.5\,mag (between ages of 4 and 14\,Gyr). In addition, the growth of an NSC is affected by a variety of factors, e.g. the gravitational environment (i.e. the shape of the host galaxy potential) or the abundance of molecular gas in the nucleus, both of which will affect the star formation rate in and around the NSC. 
 
As for the structure of NSCs, only small changes ($<\!10$\%) in $r_{\rm eff}$ are expected due to their internal dynamical evolution, because $r_{\rm eff}$ is stable over many relaxation times after the first Gyr, during which stellar mass loss dominates. Larger changes in $r_{\rm eff}$ are expected from {\it external} mechanisms such as the infall of other star clusters or {\it in situ} star formation caused by gas accretion onto the NSC.

In almost all galaxies in our sample (95\% of the cases), the light distribution of the NSCs is better described by a King- than an EFF profile. For more than half of these (58\%), the King profile with the highest concentration index (King100, i.e. $C\equiv r_t/r_c=100$) provides the lowest $\chi^2$ value. Other concentration indices are optimal for smaller sample fractions as follows: King30 for 22\%, King15 for 12\%, and King5 for 8\% of NSCs. A comparison of these statistics to other studies is not easily possible, since \cite{Seth08} adopted a fixed King model with $C=15$ for their late-type sample of NSCs, while \cite{Cote06} did not provide the concentration indices of their best-fit models for their sample of early-type nuclei. The nine nuclear clusters in the low-mass late-type dwarf galaxy sample in \cite{Georgiev09b} are also better fit with King models that have similar concentrations, namely $C=100\ (40\%),\ 30\ (20\%),\ 15\ (30\%)$ and 5 (10\%). We emphasize that for a given NSC, the distinction between models with different concentration indices may not always be definitive, since the differences in the fit quality are often small. Nevertheless, the statistics appear to indicate that generally speaking, the structure of NSCs is closer to that of UCDs \cite[$C=63$ is the median concentration of 40 UCDs in][]{Mieske13} than to that of a typical Galactic GC \cite[$C=32$ is the median concentration in][]{Harris96}. 

\subsection{Comparison to other compact stellar systems}\label{Sect: NSCs_GC_UCDs}
%%%%%%%%%%%%%%%%%%%%%%%%%
Luminosity, mass, and effective radius are fundamental properties of compact stellar systems that reflect their formation and subsequent dynamical evolution in the nuclear environment \citep[e.g.][]{Merritt09}. A comparison of these quantities between different stellar systems (e.g. massive GCs, UCDs, NSCs) thus can provide valuable information on their dynamical status and possible evolutionary connection. It has been proposed that all these different incarnations of massive, dense, and dynamically hot stellar systems share a common origin as the remnant nuclei of now defunct galaxies following their tidal stripping/dissolution. The size measurements of NSCs in our sample show a broad range, from being unresolved even for some nearby nuclei (cf Fig.\,\ref{fig: reff_dist}) to a few tens of parsecs, a regime that is comparable to the size of UCDs.
\begin{figure*}
\epsfig{file=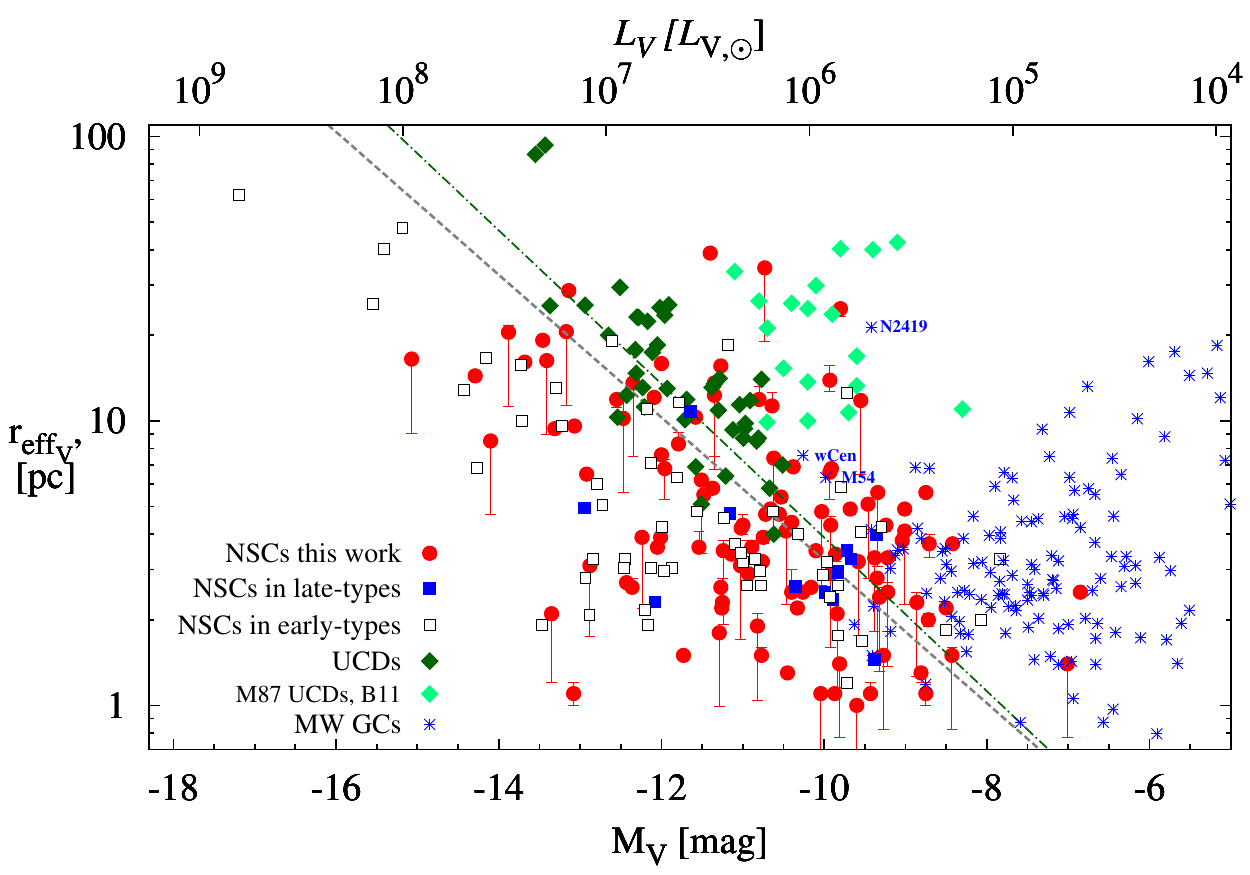, width=1\textwidth, bb=20 5 350 250}
\caption{Comparison of effective radius versus absolute luminosity ($M_V$) between the NSCs in this study (red circles) and literature data for NSCs in late-type spiral or irregular galaxies (blue squares), NSCs in early-type hosts (open squares), and UCDs (green diamonds), as referenced in \S\,\ref{Sect: NSCs_GC_UCDs}. The dashed and dash-dotted lines show, respectively, our fit to the NSC in our sample (eq.\,\ref{eq:size-mass}) and the \protect\cite{Evstigneeva08} fit to UCDs.
}\label{fig: reff Mv compare}
\end{figure*}

In Figure\,\ref{fig: reff Mv compare} we compare our measured NSC sizes to those published in other studies, in particular for early-type (dE,N) galaxies \cite[open squares]{Cote06}, and other late-type and irregular galaxies \cite[blue squares]{Seth06,Georgiev09b}. 
We also plot the $r_{\rm eff}$ data for UCDs (green diamonds) from a number of studies \citep{Evstigneeva08,Norris&Kannappan11,Misgeld11,Mieske13}, as well as those of Milky Way GCs (asterisks) from the 2010 update of the \cite{Harris06a} catalogue. 
Some of the most massive Galactic GCs are explicitly labelled. We also plot the new size measurements of \cite{Brodie11} for the UCDs around M\,87 which have been confirmed spectroscopically by \cite{Strader11}. 
Note that \citeauthor{Brodie11} adopt a fixed King30 model to measure the UCD sizes with {\sc ishape}.

The dashed line in Figure\,\ref{fig: reff Mv compare} shows again the best fit to our NSC sample from Fig.\,\ref{fig: WFPC2_reff_Mv_BVI_all}, while the dashed-dotted line marks the fit to the distribution of early-type NSCs and UCDs obtained by \citet[their eq. 6]{Evstigneeva08}. 

It is evident from Figure\,\ref{fig: reff Mv compare} that NSCs in late-type galaxies bridge the parameter space between the more compact NSCs in early-type galaxies and the more extended UCDs over most of the luminosity range covered by these systems. At a given luminosity, UCDs are about two times larger than early-type NSCs. As discussed by \cite{Evstigneeva08}, this can be understood as the effect of the strong and variable tidal truncation within the steep core of early-type nuclei, which is not present for isolated UCDs. 
This is also supported by the results of numerical experiments \citep{Bekki&Freeman03,Ideta&Makino04} which have shown that compact nuclear clusters can expand and become UCD-like systems. More specifically, the recent numerical simulations of \cite[e.g.][]{Pfeffer&Baumgardt13} demonstrate that the tidal stripping of a nucleated early-type dwarf galaxy following a close ($\leq10$\,pc) pericentre passage by a more massive galaxy 
can form stellar systems with properties that cover the entire range of observed UCD sizes and luminosities, including the faint and extended UCDs found in \cite{Brodie11}. 

\cite{Pfeffer&Baumgardt13} also show that in low-mass galaxies with a shallow potential, expansion of a stripped nucleus may not be as pronounced. Nevertheless, the fact that a few late-type NSCs in our sample overlap with those extended UCDs, suggests that even without significant expansion they may resemble UCDs after disruption of their host. Similarly, the most luminous GCs also show significant overlap with NSCs, despite the fact that they likely experienced significant fading due to both passive ageing and mass loss. Taken together this could be interpreted as support for a scenario in which UCDs and at least some massive GCs share a common origin as the former nuclei of now dissolved galaxies which were destroyed in past encounters with a more massive galaxy.  

Such a scenario also offers a natural explanation for the observed complex/multiple stellar populations in a number of massive Galactic GCs such as $\omega$Cen and M\,54.
This is because the location of the NSC in the nucleus of a host galaxy, i.e. at the bottom of a deep potential well, 
favours a number of mechanisms which are likely to result in the build-up of various generation of stars via mechanisms such as {\it in situ} star formation or the infall and merging of stellar clusters due to dynamical friction.

%%%%%%%%%%%%%%%%%%%%%%%%%
\subsection{Colours and Stellar Populations}\label{Sect: NSC stellar pops}
%%%%%%%%%%%%%%%%%%%%%%%%%

As discussed in \S\,\ref{Sect: NSC photometry},
the NSC magnitudes are computed from the best-fit {\sc ishape} model. For our analysis, we will compare the measurements to magnitudes and colours of single stellar populations (SSPs) as predicted by the 2012 update of the \cite{BC03} models. These models incorporate the latest stellar evolutionary tracks of thermally-pulsating asymptotic giant branch stars (TP-AGB) of different masses and metallicities by \cite{Marigo08}, which are especially important for clusters with 
a luminosity weighted age of 0.2\,-\,2\,Gyrs, where the contribution of TP-AGB stars is expected to be maximal.
We focus the comparison to solar or higher metallicities, because high-resolution spectroscopy of some NSCs in late-type spirals has shown that they are best described by slightly subsolar (Z$=0.015$) or higher metallicity \citep{Walcher06}.

\begin{figure*}
\epsfig{file=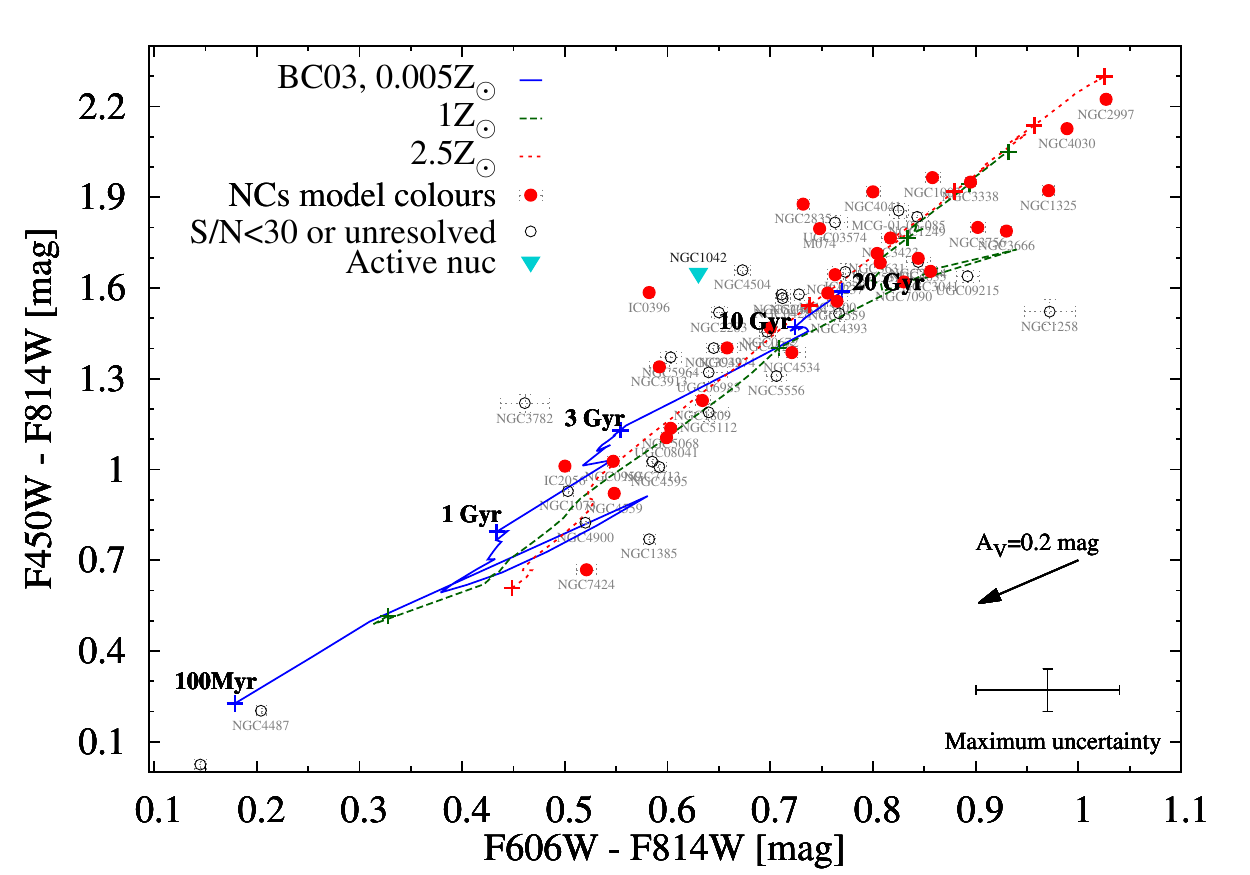, width=1\textwidth, bb=0 5 360 252}
\caption{Colour-colour diagram of all NSCs with photometry in the $F450W, F606W$ and $F814W$ filters. Red circles denote NSCs with reliable size measurements, while the open circles are for those with only upper limits for $r_{\rm eff}$, as discussed in \S\,\ref{Sect: Uncertainties}. The magnitudes in all three filters have been corrected for Galactic foreground reddening. The arrow indicates a dereddening vector of $A_V=0.2$\,mag, and the error bar indicates the photometric uncertainty as discussed in \S\,\ref{Sect: Uncertainties}. Three evolutionary tracks for SSPs \protect\citep{BC03} with different metallicities are overplotted.
}\label{fig:WFPC2_CCD}
\end{figure*}

In Figure\,\ref{fig:WFPC2_CCD}, we show the colour-colour diagram for all NSCs that have suitable data in the three most common WFPC2 filters, i.e. $F450W, F606W$ and $F814W$. Most NSCs fall close to the SSP models, which is a confirmation for the homogeneity and quality of our photometric measurements. It is important to note that any variation between NSCs in the properties of their stellar population(s), e.g. in age, metallicity, and/or extinction would cause their position in Figure\,\ref{fig:WFPC2_CCD} to be shifted \emph{along} the SSP tracks. The uncertainty in the measured colours (indicated in the lower right), is derived from the maximum photometric error of 0.07\,mag, %which is based on rather conservative assumptions 
as discussed in \S\,\ref{Sect: Uncertainties}. We therefore believe the displacement of a subset of NSCs above and to the left of the model tracks cannot fully be explained by uncertainties in the photometric measurements. Instead, their bluer $F606W - F814W$ colours are most likely indicative for excess flux in the $F606W$ filter due to $H_\alpha$ emission produced by ongoing star formation and/or weak AGN activity \cite[e.g.][]{Seth08}. 

This seems to be confirmed by the position of NGC\,1042, (blue triangle), the only object in the weak AGN comparison sample (see \S\,\ref{Sect: AGNs}) with photometry in all three bands. It clearly cannot be explained by purely stellar emission, which suggests that this diagnostic may be useful for identifying similar cases with weak AGN activity. Deep, small-aperture spectroscopy or narrow-band imaging with HST-like spatial resolution is needed to confirm the presence of AGN signatures in other NSCs 
that have colors not matching the model predictions.

\begin{figure}
\epsfig{file=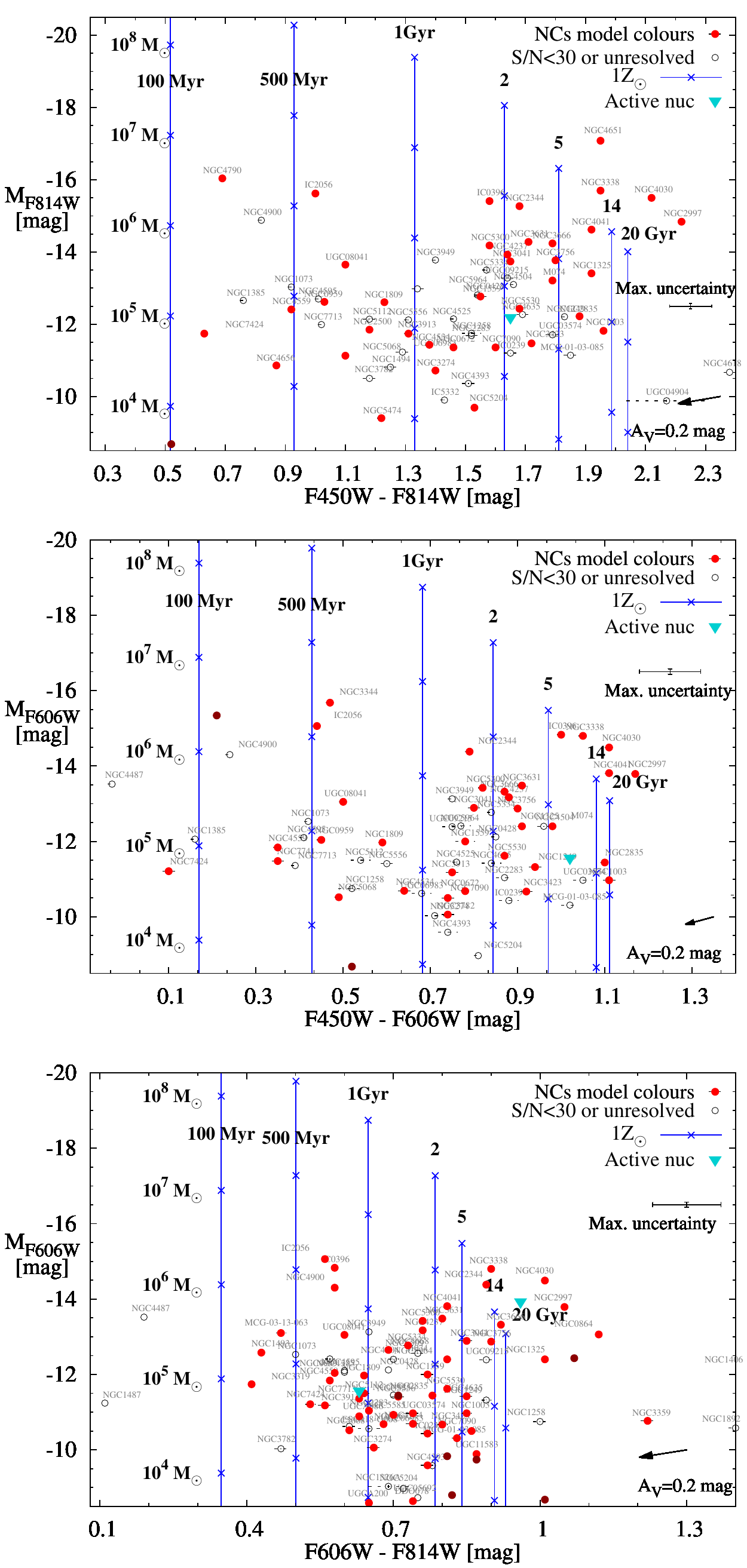, width=.5\textwidth, bb=10 5 380 750}
\caption{Colour-magnitude diagrams for NSCs observed in the three most frequently used WFPC2 filters. All three colours are plotted against absolute magnitude. The measurements are compared to solar me\-ta\-llicity isochrones (solid blue lines), cro\-sses and diamonds along the isochrones indicate the stellar masses corresponding to the respective luminosity, according to the \protect\cite{BC03} SSP model $M/L$s. Solid (red) circles denote NSCs with reliable size measurements, while open circles 
are for NSCs with upper $r_{\rm eff}$, as discussed in \S\,\ref{Sect: Uncertainties}. All magnitudes have been corrected for Galactic foreground reddening. The arrow indicates a dereddening vector of $A_V\!=\!0.2$\,mag, and the error bar indicates the maximum photometric uncertainty (cf. \S\,\ref{Sect: Uncertainties}). 
}
\label{fig:NSCs CMD}
\end{figure}

In order to further gauge the range of stellar population ages covered by NSCs, we show in Figure\,\ref{fig:NSCs CMD} colour-magnitude diagrams, together with a number of isochrones for solar and super-solar metallicity. Along each isochrone, we indicate the corresponding cluster masses derived from the $M/L$-ratios predicted by the SSP model for the respective age. 

It is evident from Figure\,\ref{fig:NSCs CMD} that the NSCs in our sample span a wide range in age. About one third of NSCs have blue colours consistent with luminosity-weighted ages younger than 1-2\,Gyrs, assuming solar metallicity. Given that a young stellar population (of, e.g., 500\,Myr) outshines an older population (of, e.g., 14\,Gyr) by $M_V\simeq2$\,mag at the same metallicity and mass, even a small fraction (10\% in mass) of young stars will significantly bias the integrated luminosity and colours towards an age younger than that of the older stellar population(s) dominating the cluster mass. Therefore, it is not possible to use Figure\,\ref{fig:NSCs CMD} to determine the time of NSC formation, i.e. the age of the {\it oldest} stellar population within each NSC. 

Nevertheless, it is evident that recent star formation is ubiquitous in the nuclei of late-type spiral galaxies. This result is in agreement with previous spectroscopic studies of smaller NSC samples \cite[e.g.][]{Walcher06,Rossa06}. 

%%%%%%%%%%%%%%%%%%%
\subsection{Double nuclei and nuclear disks}\label{Sect: multiple NSCs}
%%%%%%%%%%%%%%%%%%%
In this section, we comment on those galaxies that were rejected from the NSC catalogue for one or more of the reasons described in \S\,\ref{Sect:Gal-samp}. About 15\% (47 galaxies) of the candidate late-type disks have nuclear morphologies that are too complex to derive reliable NSC properties with our automated methods. While a more detailed analysis of these nuclei is beyond the scope of this paper, we will highlight a few examples which illustrate the two main processes suggested to drive the evolution and growth of an NSC, namely star cluster merging and {\it in situ} star formation triggered by gas infall into the nucleus.

The formation of an NSC via cluster merging in the galaxy nucleus \cite[e.g.][]{Tremaine75} is a natural consequence of dynamical friction which can be an efficient mechanism for the orbital decay of star clusters and their subsequent migration to the galaxy nucleus \citep{Chandrasekhar43,Tremaine75,Agarwal&Milosavljevic11}.
Numerical simulations have demonstrated that this process can form clusters with masses and structural properties comparable to those of observed NSCs \cite[e.g.][]{Capuzzo-Dolcetta93,Bekki04,Capuzzo-Dolcetta&Miocchi08,Antonini13}, including those in disk galaxies \citep{Bekki10,Hartmann11}.

\begin{figure}
\includegraphics[scale=.31]{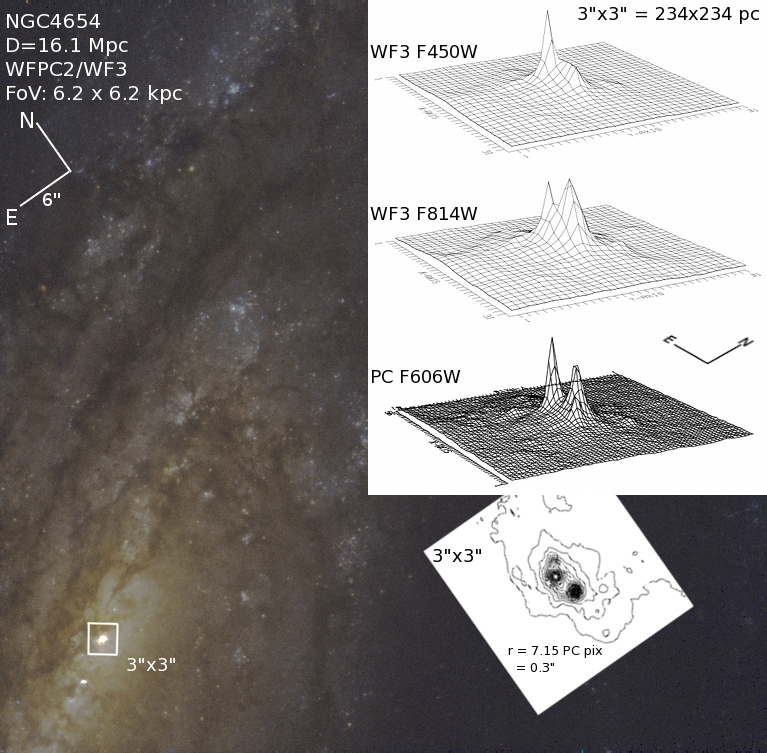}
\caption{An example of a double nucleus in NGC\,4654. The colour composite image is constructed from exposures in the $F450W$ and $F814W$ WFPC2/WF3 filters. The size of the nuclear region highlighted in the surface and contour plots is $3\arcsec\times3\arcsec=234\times234$\,pc. 
The separation between the two sources evident in the bottom contour plot is 7.15 PC pixels = 27.9\,pc.
}
\label{fig: NGC4654_BI_3_mos}
\end{figure}

A prominent example for the presence of two clusters in the galaxy nucleus that are likely to merge within a few crossing times is presented in Figure\,\ref{fig: NGC4654_BI_3_mos}. It shows a color-composite image of NGC\,4654, together with surface and contour plots of its nuclear region (marked with a white square). This galaxy is located in the Virgo cluster, at a distance of 16.1\,Mpc, and has a luminosity of $M_V=-20.1$\,mag, comparable to the Milky Way. It is a late-type spiral ($t=5.6$, SBc) with an inclination of $i=60^\circ$. The projected separation between the two clusters is 7.15 pixels on the PC detector, i.e. $0.36\arcsec$ or 27.9 pc. This separation is about 10 times larger than the typical $r_{\rm eff}$ of an NSC (cf Fig.\,\ref{fig: WFPC2_reff_hist_BVI}). 

The physical proximity of the two clusters prevents an accurate {\sc ishape} fit to their structure and luminosity with the automated approach used for the current study, which is why this and other similar cases have been excluded from the NSC catalogue presented here. Nevertheless, we have performed simple aperture photometry within a two pixel radius which shows that the relative luminosities of the two clusters vary strongly between the different filters, suggesting that they have stellar populations of rather different ages.

More specifically, both clusters have about the same luminosity in the $F606W$ image, with the Eastern cluster being brighter by about 0.1 mag. However, the Eastern cluster is by far the brighter of the two in the $F450W$ exposure, yet it is actually fainter in $F814W$. The blue colour of the Eastern cluster suggests that it is likely dominated by young stars with ages less than a few hundred Myrs. Comparison to SSP models implies that it is at least an order of magnitude less massive than its Western counterpart, which has a mass of about $10^6 M_\odot$. 

Assuming that the two clusters are gravitationally bound to each other, we can estimate roughly their kinematics using the measured separation and their approximate masses. 
The measured separation between the two clusters of $r = 28$\,pc implies an orbital period of $P = 9.3\times10^7\ a^{3/2}\ (m1 + m2)^{−1/2}\simeq50$\,Myr, where $a=r/2\simeq14$\,pc and $m1 + m2 = 1.1^6 M_\odot$ are the orbital semi-major axis in pc and the sum of the cluster masses, respectively. This should be compared to the Roche radius for a binary system,\\
$R_R = a\ (\ 0.38 + 0.2\ \log(m1/m2)\ )^{1/2} = 11.4$\,pc, \citep{Makino91,Paczynski71}. Since the projected cluster separation is comparable to the calculated Roche radius, numerical simulations predict that within few orbital periods equal mass binary clusters would merge \citep{Sugimoto&Makino89}. It is therefore very likely that these two clusters in the nucleus of NGC\,4654 will merge within 0.5\,Gyr. Note, however, that some studies have indicated that an apparent double nucleus can be supported in a dynamically stable configuration over a long time in the presence of a SMBH, e.g. in M31 \citep{Kormendy&Bender99}, NGC\,4486B \citep{Lauer96}, and VCC 128 \citep{Debattista06}.

The assumption that both clusters are gravitationally bound can, in principle, be tested by comparing their orbital velocity to a measurement of the velocity dispersion of the system, i.e. within an aperture of $<1\arcsec$. If both clusters indeed form a bound pair, their orbital velocity would be $V \simeq\sqrt{G\, m_1/r}=12.4$\,km/s, for an assumed circular orbit, and an NSC mass of $m_1 = 10^6M_\odot$. 

\begin{figure}
\includegraphics[scale=.32]{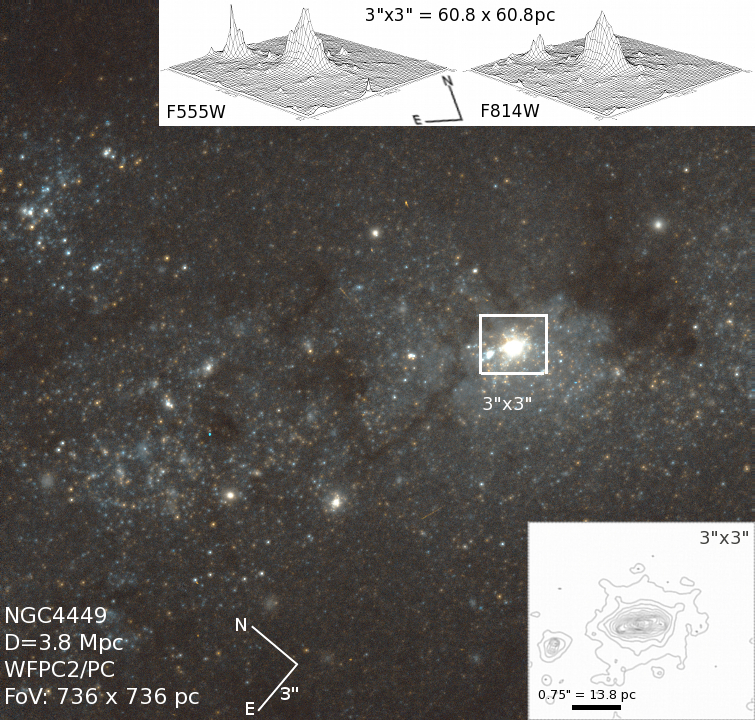}
\caption{An example of a nuclear disk in NGC\,4449. The colour composite image from the WFPC2/PC detector in the $F555W$ and $F814W$ filters. The size of the inlays is $3\arcsec\times3\arcsec=60.8\times60.8$\,pc. In the bottom contour plot, we also show the scale size of the nuclear disk.}
\label{fig: NGC4449_VI_1_mos}
\end{figure}

Another viable channel for NSC formation is NSC growth via \emph{in situ} star formation triggered by gas accretion, as proposed by \cite[e.g.][]{Bekki07,Seth08b,Neumayer11,DeLorenzi11}. This process is especially viable in spiral galaxies, which normally harbour large amounts of molecular gas that can be funneled towards the nucleus by a number of dynamical processes \citep{Kormendy13}.

To demonstrate the importance of this mechanism in present-day galaxies, we present an example of a circum-nuclear disk in Figure\,\ref{fig: NGC4449_VI_1_mos} which shows a colour-composite image of NGC\,4449, a nearby, isolated, barred Magellanic-type irregular galaxy (IBm, $t=9.8$) whose inclination is estimated to be $i=63.5^\circ$. The contour plot shows a narrow, elongated emission region that is about 14\,pc long by less than 2\,pc wide.

Earlier spectroscopic studies of the nucleus of NGC\,4449 have found a number of emission lines which indicate that it is the site of ongoing star formation \citep{Boeker01,Hunter05}. It is therefore plausible that this structure is a circum-nuclear disk which is in the process of forming stars, and that these will contribute to the growth and rejuvenation of the NSC in NGC\,4449. This disk has the same orientation as the host galaxy disk (PA\,$=46.5^\circ$), which is in line with observation of other young nuclei \citep[e.g.][]{Seth06,Seth08b}.

In summary, these two examples provide evidence that {\it both} of the two proposed NSC growth mechanisms do indeed occur in the present-day universe. Their relative importance {\it today} depends on the current properties of the host galaxy, e.g. the size of the available gas reservoir within the galactic disk or the shape of its gravitational potential which determines whether massive star clusters that formed elsewhere in the central region are able to survive long enough to migrate inwards, and to form an NSC or to merge with a pre-existing NSC.

\subsection{Active nuclei in the sample}\label{Sect: AGNs}

Emission from an active Galactic Nucleus (AGN) is likely to affect its colour and apparent structure. Because it will appear as an additional point source ``on top of'' the NSC proper, AGN emission will make it difficult to measure the intrinsic size of the NSC. For this reason, we excluded all strong AGNs when constructing our galaxy sample.

However, because the study of weak AGN in bulge-less disk galaxies is an important topic in its own right, identifying an efficient method to search for these rare objects \citep[]{Satyapal09} is a potential benefit of our study. In addition, a systematic comparison of the structural and photometric properties of NSCs with and without weak AGNs has the potential to provide additional constraints on the mechanisms powering such nuclei. For example, the presence of H$_\alpha$ emission could modify the structure and flux in the $F606W$ filter (cf. Sect.\,\ref{Sect: NSC stellar pops}).
 
For these reasons, we kept the small subsample of weak active nuclei in our sample (see \S\,\ref{Sect:Gal-samp}), in order to compare their properties to those of quiescent NSCs. More specifically, the AGN sub-sample consists of the following eight galaxies, which are classified as a Seyfert and/or H2 class in the literature: NGC\,1042 \citep{Shields08}, NGC\,4395 \citep{Filippenko&Sargent89}, NGC\,1058, NGC\,1566, NGC\,3259, NGC\,4639, NGC\,4700, NGC\,5427 \citep{Veron-Cetti06}.

\begin{figure}
\includegraphics[scale=.32]{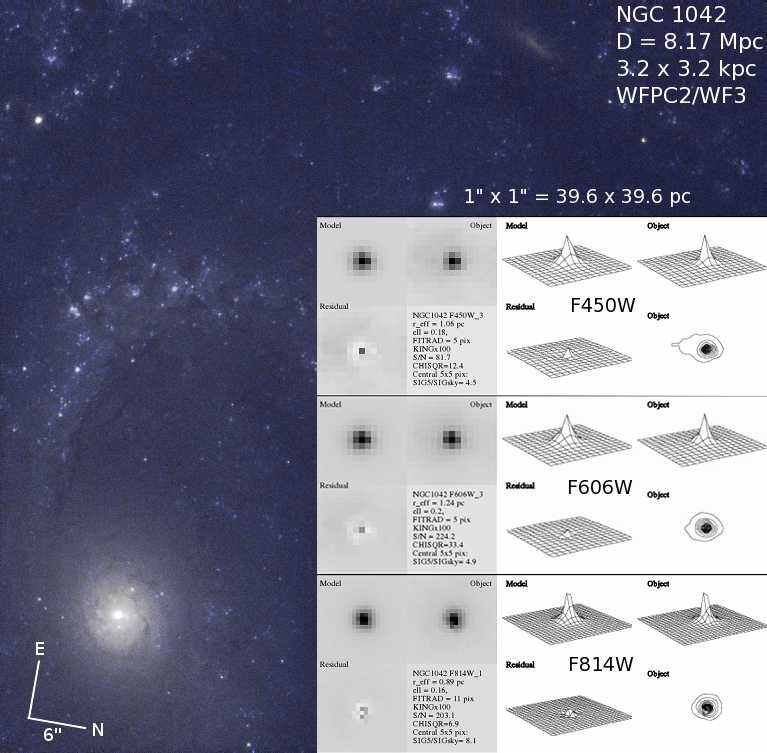}
\includegraphics[scale=.32]{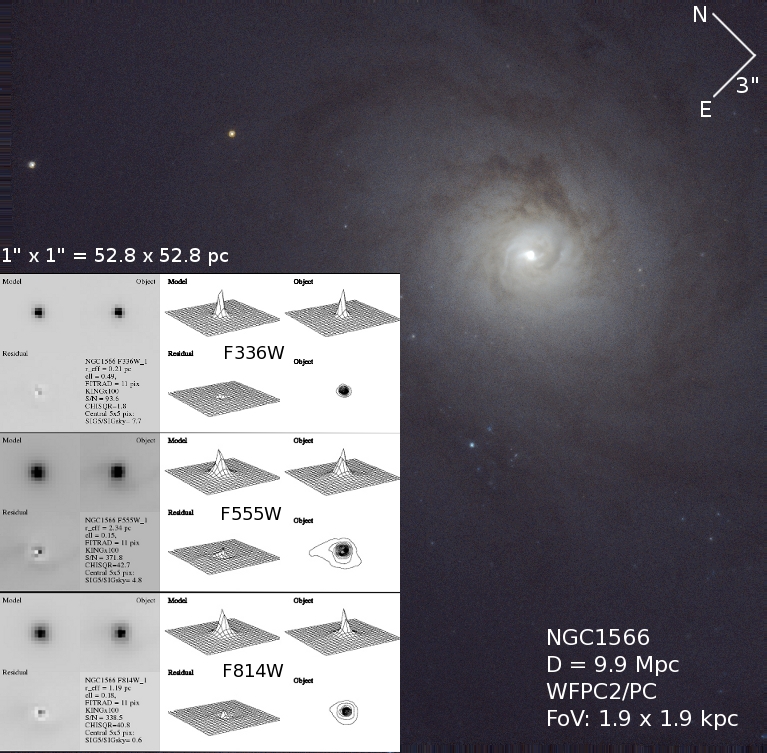}
\caption{Two examples for NSCs with a weak AGN: NGC\,1042 (top) and the Seyfert 1.5 galaxy NGC\,1566 (bottom). The colour composite image of NGC\,1042 is made from WFPC2/WF3 $F450W$ and $F606W$ images, while for NGC\,1566, we used the WFPC2/PC1 $F555W$ and $F814W$ exposures. The {\sc ishape} fitting results show unresolved residual emission, indicating the presence of an additional point source. 
}\label{fig: NGC1042_BV_3_mos}
\end{figure}

Suitable HST/WFPC2 images in more than one filter are available only for NGC\,1042 ($F450W, F606W, F814W$), NGC\,1058 ($F450W$ and $F606W$) and NGC\,1566 ($F336W, F555W$ and $F814W$). In Figure\,\ref{fig: NGC1042_BV_3_mos}, we show color composite images of NGC\,1042 and NGC\,1566. NGC\,1042 is a weak AGN \citep{Shields08} which has an undefined {\tt agnclass} parameter in HyperLEDA. NGC\,1566 is classified as a Seyfert 1.5, i.e. it has rather strong emission from H$\alpha$ and H$\beta$. 

Not surprisingly, we find that NSCs with an AGN generally appear more compact than ``normal'' NSCs at a given luminosity, as shown in Figure\,\ref{fig: WFPC2_reff_Mv_BVI_all}. Their optical colours are expected to be affected by the AGN, such that they depart from the predictions of SSP models. As seen from the colour-colour diagram in Fig.\,\ref{fig:WFPC2_CCD}, this is indeed the case for NGC\,1042, the only object that we can test this prediction on. This NSC is clearly bluer than the SSP models which can not be explained by any plausible mix of age, metallicity or extinction. In addition, the residual images in Figure\,\ref{fig: NGC1042_BV_3_mos} show unresolved extra emission in the central few pixels at the 6-7\% level, which most likely stems from the AGN. 
We conclude that the combination of non-stellar colours and point-like residual emission in the {\sc ishape} fits promises to be a powerful tool to identify weak AGNs.

\section{Summary and Conclusions}\label{Sect:summary}

Understanding the formation and evolution of nuclear star clusters (NSCs) promises fundamental insight into their relation to other massive compact stellar systems. This is because systems such as ultra compact dwarf (UCD) galaxies or massive globular clusters (GCs) harboring multiple stellar populations possibly originate as the former nuclei of now defunct satellite galaxies. On the other hand, NSCs often coexist with central black holes at the low-mass end of the SMBH mass range (around $\simeq 10^6 M_{\odot}$). It is actively debated what the role of NSCs is in the growth of such black holes and the fuelling of energetically weak ``mini-AGNs''. To address these questions, it is important to provide reliable measurements of the stellar populations properties of NSCs (age, metallicity, mass), as well as of their structural parameters for as many NSCs as possible, to provide observational constraints to the growing body of theoretical work addressing the above topics. 

The NSC catalogue presented in this work is the first step in this direction. It provides the largest and most homogeneously measured set of structural and photometric properties of nuclear star clusters in late-type spiral galaxies, derived from HST/WFPC2 archival imaging.

We have searched the HST legacy archive for all late-type spirals within 40 Mpc (see Sect.\,\ref{Sect:Gal-samp}, Table\,\ref{Table:Galaxy sample}) that were observed with WFPC2. We have identified 323 such galaxies with suitable images of their nuclear region. More than two thirds (228/323) of these show an unambiguous NSC. We have used a state-of-the-art customized PSF-fitting technique to derive robust measurements of their effective radii and luminosities. 

For the PSF-fitting of each NSC, detailed in Section\,\ref{Sect: Measuring NSC sizes}, we employ {\sc TinyTim} PSF models tailored to the pixel location on the respective WFPC2 detector. We use the {\sc ishape} software \citep{Larsen99} to perform a $\chi^2$ minimisation fitting between the observed NSC profile and the PSF model convolved with an analytical model cluster profile. During this step, the fitting radius is iteratively adapted to the size of the NSC, minimizing the impact of circum-nuclear structures. We use the best-fit {\sc ishape} model to derive both the structural parameters ($r_{\rm eff}$, $PA$, and $\epsilon$) and the photometry of each NSC. For the latter, we use the latest \citep{Dolphin09} transformations, together with the measured NSC color, if available (see \ref{Sect: NSC photometry}). 

The complete catalogue of all measured structural and photometric properties of the 228 NSCs analysed in the various WFPC2 filters is provided in the online version of this paper. Table\,\ref{Table:Galaxy sample} contains the basic properties of the sample galaxies. Effective radius measurements for different filters and best fit analytical profiles are provided in Table\,\ref{Table:reff}, 
while the ellipticities and position angles of all NSCs are provided in Table\,\ref{Table:ellpa}.
Calibrated and foreground reddening corrected NSC model magnitudes in the WFPC2 magnitude system are listed in Table\,\ref{Table: F300-814W_mag}, and their magnitudes in the Johnson/Cousins system in Table\,\ref{Table: UBVI_mag}. We caution that the latter magnitudes should be used with caution, because their accuracies depend on the available colour information for the respective NSC.

Our main results can be summarized as follows:
\begin{itemize}
\item{ {\bf Sizes and structure:} we find that 
the measured sizes of NSCs in late-type spiral galaxies cover a wide range. Most NSCs have $r_{\rm eff}$ of a few pc, typical for Milky Way GCs. However, the $r_{\rm eff}$ distribution includes NSCs as large as a few tens of pc, i.e. comparable to some UCDs. There is tentative evidence for a smaller mean NSC size at bluer wavelengths, possibly caused by the presence of a weak AGN and/or a young stellar population that is more concentrated than the bulk of the NSC stars. On the other hand, some NSCs appear to be about 30\% {\it larger} when observed in the $F606W$ filter, compared to measurements in other filters. We discuss that this could be due to H$_\alpha$ line emission from (circum-)nuclear star formation. The NSCs in our late-type galaxy sample fall into the size-luminosity parameter space between early-type nuclei, UCDs, and massive GCs, which we interpret as support for the formation of these dense stellar systems from remnant nuclei of disrupted satellite galaxies (see \S\,\ref{Sect: NSCs_GC_UCDs}).
The majority of the NSCs in our sample are best fit with King profiles with a high-concentration index ($C\equiv r_t /r_c  = 100$), which makes them structurally similar to  UCDs (see \S\,\ref{Sect: Size Distribution}). } 

\item {\bf Stellar populations:}
the colour-colour and colour-magnitude diagrams (Fig.\,\ref{fig:WFPC2_CCD} and \ref{fig:NSCs CMD}) show that NSCs span a wide range in age and/or metallicity of their stellar populations. This agrees with previous studies which have suggested that NSCs are likely to experience recurring star formation events and/or accretion of other stellar clusters. 
A comparison to SSP models shows that NSCs also span a wide range in mass, from a few times $10^4$ to a few times $10^8\cal{M}_\odot$. Unfortunately, stellar mass estimates from optical photometry alone are rather uncertain due to the strong variation in $M/L$ as a function of age. As discussed in \S\,\ref{Sect: NSC stellar pops}, a small contribution from a young (e.g. 0.5 - 1\,Gyr) stellar population can be as luminous as a (much more massive) older (e.g. 10\,Gyr) stellar population, thus strongly biasing the derived cluster age.
Combination of this catalogue with UV- and/or near-infrared data would provide much more robust mass, metallicity, and age estimates for NSCs.

\item {\bf Double nuclei and nuclear disks:}
We find a number of galaxies hosting double nuclei, i.e. two star clusters with comparable luminosity which are separated by only a few tens of parsecs. 
We regard these cases as plausible examples for the ongoing process of clusters merging in the galaxy nucleus (\S\,\ref{Sect: multiple NSCs}).
We also find examples for small-scale circum-nuclear disk (aligned with host galaxy disk), which we interpret as evidence for NSC growth via gas accretion \citep{Seth06}. A systematic search for such morphological features in HST images can provide important constraints on each of the proposed NSC formation channels, and a statistical analysis of their frequency will be the topic of a follow-up paper. 

\item {\bf Active nuclei:}
We also analysed a small comparison sample of weak AGNs. In some of these cases, we find faint unresolved nuclear emission in the residuals of the best-fit cluster model which are most likely caused by the AGN. The only AGN with complete colour information (NGC\,1042) deviates from SSP model predictions, suggesting that the AGN emission significantly affects the NSC colour.
We therefore argue that such PSF-fitting techniques can be used to search for so far undetected nuclear activity, or at least to define promising target samples for spectroscopic searches for similar weak AGNs.

\end{itemize}

\section*{Acknowledgments}
We are grateful to the referee, Anil Seth, for constructive comments and suggestions, in particular on the discussion of NSC position angles and their alignment with the host galaxy disk. We also would like to thank Dr. S\o{}ren Larsen for valuable discussions. IG acknowledges support from the Euroepan Space Agency via the International Research Fellowship Programme. We also acknowledge the use of the HyperLeda database (${\rm http://leda.univ-lyon1.fr)}$. This research has made use of the NASA/IPAC Extragalactic Database (NED) which is operated by the Jet Propulsion Laboratory, California Institute of Technology, under contract with the National Aeronautics and Space Administration.

\clearpage
\newpage

\begin{landscape}
\begin{deluxetable}{lrrlllllllllll}
\tabletypesize{\footnotesize}
\tablecolumns{14}
\tablewidth{0pc} %%% <--- This is important!!! Otherwise it wont compile!!!!
\tablecaption{
Main properties of the galaxy sample with measured NSC properties. \\
(All 228 galaxies are listed in the online version of the table).
\label{Table:Galaxy sample}
}
\tablehead{

\colhead{Galaxy} &
\colhead{RA} &
\colhead{DEC} &
\colhead{$m-M$} &
\colhead{E($B-V$)} &
\colhead{$B$} &
\colhead{$B-V$} &
\colhead{$I$} &
\colhead{R$_{25}$} &
\colhead{$\epsilon$} &
\colhead{PA} &
\colhead{Incl.} &
\colhead{type} &
\colhead{t} \\
\colhead{} &
\colhead{hh:mm:ss} &
\colhead{dd:mm:ss} &
\colhead{mag} &
\colhead{mag} &
\colhead{mag} &
\colhead{mag} &
\colhead{mag} &
\colhead{kpc} &
\colhead{} &
\colhead{deg} &
\colhead{deg} &
\colhead{} &
\colhead{} \\
\colhead{(1)} &
\colhead{(2)} &
\colhead{(3)} &
\colhead{(4)} &
\colhead{(5)} &
\colhead{(6)} &
\colhead{(7)} &
\colhead{(8)} &
\colhead{(9)} &
\colhead{(10)} &
\colhead{(11)} &
\colhead{(12)} &
\colhead{(13)} &
\colhead{(14)}
}
\startdata

DDO078	&	10:26:27.78	&	67:39:25.1	&	27.82	&	0.018	&	15.8	&	\nodata	&	\nodata	&	1.063	&	0.00	&	\nodata	&	0.	&	I	&	10.	\\
IC4710	&	18:28:37.95	&	-66:58:56.1	&	29.75	&	0.079	&	12.51	&	0.57	&	11.19	&	4.494	&	0.15	&	\nodata	&	34.9	&	Sm	&	8.9	\\
NGC1258	&	3:14:05.50	&	-21:46:27.3	&	32.28	&	0.022	&	13.88	&	\nodata	&	12.35	&	5.870	&	0.26	&	20.5	&	43.7	&	SABc	&	5.7	\\
NGC3319	&	10:39:09.47	&	41:41:12.5	&	30.7	&	0.013	&	11.77	&	0.41	&	11.46	&	7.289	&	0.51	&	36.	&	62.7	&	SBc	&	5.9	\\
NGC5334	&	13:52:54.44	&	-1:06:52.4	&	32.78	&	0.041	&	12.97	&	\nodata	&	12.19	&	17.729	&	0.28	&	18.2	&	44.8	&	Sc	&	5.2	\\

\nodata&\nodata&\nodata&\nodata&\nodata&\nodata&\nodata&\nodata&\nodata&\nodata&\nodata&\nodata&\nodata&\nodata\\

\enddata

\vspace{-.5cm}

\tablecomments{The values for all columns are taken from HyperLEDA, except for columns 4 and 5, which are taken from NED. More specifically, the distance modulus $m-M$ in Column 4 is the median value in NED. If the latter is not available, we adopt the redshift-derived distance modulus, modz, from HyperLEDA. 
}

\end{deluxetable}

\begin{deluxetable}{lrrllllllllllll}
\tabletypesize{\footnotesize}
\tablecolumns{15}
\tablewidth{0pc} %%% <--- This is important!!! Otherwise it wont compile!!!!
\tablecaption{
Main properties of the galaxy sample without an identifiable NSC. \\
(All 95 galaxies are listed in the online version of the table).
\label{Table:Excluded galaxies}
}
\tablehead{

\colhead{Galaxy} &
\colhead{RA} &
\colhead{DEC} &
\colhead{$m-M$} &
\colhead{E($B-V$)} &
\colhead{$B$} &
\colhead{$B-V$} &
\colhead{$I$} &
\colhead{R$_{25}$} &
\colhead{$\epsilon$} &
\colhead{PA} &
\colhead{Incl.} &
\colhead{type} &
\colhead{t} &
\colhead{Comment} \\
\colhead{} &
\colhead{hh:mm:ss} &
\colhead{dd:mm:ss} &
\colhead{mag} &
\colhead{mag} &
\colhead{mag} &
\colhead{mag} &
\colhead{mag} &
\colhead{kpc} &
\colhead{} &
\colhead{deg} &
\colhead{deg} &
\colhead{} &
\colhead{} &
\colhead{} \\
\colhead{(1)} &
\colhead{(2)} &
\colhead{(3)} &
\colhead{(4)} &
\colhead{(5)} &
\colhead{(6)} &
\colhead{(7)} &
\colhead{(8)} &
\colhead{(9)} &
\colhead{(10)} &
\colhead{(11)} &
\colhead{(12)} &
\colhead{(13)} &
\colhead{(14)} &
\colhead{(15)}
}
\startdata

ESO257-G017	&	7:27:33.12	&	-45:41:04.10	&	30.26	&	0.128	&	16.75	&	\nodata	&	\nodata	&	732	&	0.17	&	71.8	&	37.3	&	SBm	&	9.	&	$F814W_1$ NoNSC	\\
ESO269-G037	&	13:03:33.19	&	-46:35:12.70	&	27.63	&	0.117	&	16.44	&	\nodata	&	\nodata	&	308	&	0.58	&	132.1	&	90.	&	IAB	&	10.	&	$F606W_3$ NoNSC $F814W_3$	\\
ESO317-G020	&	10:23:07.62	&	-42:14:14.71	&	32.55	&	0.096	&	13.21	&	\nodata	&	11.7	&	7289	&	0.09	&	\nodata	&	24.9	&	Sc	&	4.5	&	$F606W_1$ Extended/Complex	\\
IC0342		&	3:46:49.30	&	68:06:04.99	&	27.58	&	0.494	&	9.67	&	\nodata	&	\nodata	&	9521	&	0.05	&	\nodata	&	18.5	&	SABc	&	6.	&	$F606W_1$ Saturated $F555W_1$ $F675W_3$	\\
IC1613		&	1:04:47.78	&	2:07:03.83	&	24.31	&	0.022	&	10.	&	0.67	&	\nodata	&	1926	&	0.07	&	\nodata	&	22.9	&	I	&	9.9	&	$F814W_2$ NoNSC $F555W_2$ $F439W_2$	\\
IC4441		&	14:30:18.00	&	-43:33:39.63	&	30.68	&	0.147	&	14.96	&	\nodata	&	13.26	&	2231	&	0.53	&	43.1	&	65.1	&	Sc	&	4.5	&	$F300W_1$ low-$S/N$	\\

\nodata&\nodata&\nodata&\nodata&\nodata&\nodata&\nodata&\nodata&\nodata&\nodata&\nodata&\nodata&\nodata&\nodata&\nodata\\

\enddata

\vspace{-0.5cm}

\tablecomments{Columns (1)-(14) are the same as in Table\,\ref{Table:Galaxy sample}. The last column (15) identifies the reason why the galaxy was unsuitable for an NSC measurement, together with the filter and (with subscript) the WFPC2 detector of the exposure(s). 
}

\end{deluxetable}

\end{landscape}
\begin{landscape}
\begin{deluxetable}{lp{1cm}p{.3cm}p{.3cm}p{1cm}p{.3cm}p{.3cm}p{1cm}p{.3cm}p{.3cm}p{1cm}p{.3cm}p{.3cm}p{1cm}p{.3cm}p{.3cm}}
\tabletypesize{\footnotesize}
%\rotate % For landscape mode
%\setlength{\tabcolsep}{0.15cm}
\tablecolumns{16}
\tablewidth{0pc} %%% <--- This is important!!! Otherwise it wont compile!!!!
\tablecaption{
Effective radii measurements of nuclear star clusters. 
(All 228 NSCs are listed in the online version of the table, together with additional measurements in the filters F300W, F336W, F380W, F675W if available.)
\label{Table:reff}
}
\tablehead{

\colhead{} &
\multicolumn{3}{c}{  F606W } &
\multicolumn{3}{c}{  F814W } &
\multicolumn{3}{c}{  F450W } &
\multicolumn{3}{c}{  F555W } &
\multicolumn{3}{c}{  F439W } \\
\colhead{OBJECT} &
\colhead{$r_{\rm eff}$} &
\colhead{Profile      } &
\colhead{$S/N$	      } &
\colhead{$r_{\rm eff}$} &
\colhead{Profile      } &
\colhead{$S/N$	      } &
\colhead{$r_{\rm eff}$} &
\colhead{Profile      } &
\colhead{$S/N$	      } &
\colhead{$r_{\rm eff}$} &
\colhead{Profile      } &
\colhead{$S/N$	      } &
\colhead{$r_{\rm eff}$} &
\colhead{Profile      } &
\colhead{$S/N$	      } \\
\colhead{} &
\colhead{pc} &
\colhead{} &
\colhead{} &
\colhead{pc} &
\colhead{} &
\colhead{} &
\colhead{pc} &
\colhead{} &
\colhead{} &
\colhead{pc} &
\colhead{} &
\colhead{} &
\colhead{pc} &
\colhead{} &
\colhead{} \\
\colhead{(1)} &
\colhead{(2)} &
\colhead{(3)} &
\colhead{(4)} &
\colhead{(5)} &
\colhead{(6)} &
\colhead{(7)} &
\colhead{(8)} &
\colhead{(9)} &
\colhead{(10)} &
\colhead{(11)} &
\colhead{(12)} &
\colhead{(13)} &
\colhead{(14)} &
\colhead{(15)} &
\colhead{(16)}
}
\startdata

DDO078		&	$3.7_{-0.0}^{+0.1}$	&	K15$_3$	&	323.	&	$3.5_{-0.0}^{+0.0}$	&	K15$_3$	&	240.2	&	\nodata			&	\nodata	&	\nodata	&	\nodata			&	\nodata	&	\nodata	&	\nodata			&	\nodata	&	\nodata	\\
IC4710		&	\nodata			&	\nodata	&	\nodata	&	$1.0_{-0.0}^{+0.3}$	&	K100$_3$&	138.	&	\nodata			&	\nodata	&	\nodata	&	$0.8_{-0.0}^{+0.0}$	&	K100$_3$&	132.7	&	$0.9_{-0.0}^{+0.0}$	&	K100$_3$&	89.5   	\\
NGC1258		&	$_<3.3_{-0.1}^{+0.0}$	&	K100$_1$&	29.9	&	$_<4.4_{-0.8}^{+1.0}$	&	K100$_3$&	24.4	&	$_<6.8_{-6.8}^{+1.2}$	&	K100$_3$&	12.1	&	\nodata			&	\nodata	&	\nodata	&	\nodata			&	\nodata	&	\nodata	\\
NGC3319		&	$4.7_{-0.1}^{+0.1}$	&	K5$_2$	&	94.2	&	$9.4_{-0.0}^{+0.0}$	&	K5$_4$	&	267.3	&	\nodata			&	\nodata	&	\nodata	&	$9.1_{-0.0}^{+0.1}$	&	K5$_4$	&	260.7	&	\nodata			&	\nodata	&	\nodata	\\
NGC5334		&	$11.9_{-0.7}^{+0.4}$	&	K15$_3$	&	68.9	&	$14.5_{-0.6}^{+0.4}$	&	K15$_3$	&	55.1	&	$_<10.7_{-1.1}^{+0.9}$	&	K15$_3$	&	26.5	&	$9.8_{-0.3}^{+0.4}$	&	K15$_3$	&	59.6	&	\nodata			&	\nodata	&	\nodata \\

\nodata &\nodata &\nodata &\nodata &\nodata &\nodata &\nodata &\nodata &\nodata &\nodata &\nodata &\nodata &\nodata &\nodata &\nodata &\nodata \\

\enddata

\vspace{-.5cm}
\tablecomments{For each filter, we list the effective radius, the best-fitting {\sc ishape} profile, and the signal-to-noise of the respective exposure. 
The $r_{\rm eff}$ is given in pc, calculated using the distance modulus $(m-M)$ in Column 4 of Table\,\ref{Table:Galaxy sample}. The model profiles are abbreviated as K for King and E for EFF.
The subscripts in the profile column indicate the WFPC2 detector of the measurement.}

\end{deluxetable}

\begin{deluxetable}{llllllllllllll}
\tabletypesize{\footnotesize}
%\rotate % For landscape mode
%\setlength{\tabcolsep}{0.15cm}
\tablecolumns{11}
\tablewidth{0pc} %%% <--- This is important!!! Otherwise it wont compile!!!!
\tablecaption{
Ellipticities and position angles of the NSCs in our sample.
(All 228 NSCs are listed in the online version of the table, together with measurements in the F300W, F336W, F380W, F675W filters, if available.)
\label{Table:ellpa}
}
\tablehead{

\colhead{} &
\multicolumn{2}{c}{  F606W } &
\multicolumn{2}{c}{  F814W } &
\multicolumn{2}{c}{  F450W } &
\multicolumn{2}{c}{  F555W } &
\multicolumn{2}{c}{  F439W } \\
\colhead{ID} &
\colhead{$\epsilon$} &
\colhead{PA} &
\colhead{$\epsilon$} &
\colhead{PA} &
\colhead{$\epsilon$} &
\colhead{PA} &
\colhead{$\epsilon$} &
\colhead{PA} &
\colhead{$\epsilon$} &
\colhead{PA} \\
\colhead{} &
\colhead{} &
\colhead{[deg]} &
\colhead{} &
\colhead{[deg]} &
\colhead{} &
\colhead{[deg]} &
\colhead{} &
\colhead{[deg]} &
\colhead{} &
\colhead{[deg]} \\
\colhead{(1)} &
\colhead{(2)} &
\colhead{(3)} &
\colhead{(4)} &
\colhead{(5)} &
\colhead{(6)} &
\colhead{(7)} &
\colhead{(8)} &
\colhead{(9)} &
\colhead{(10)} &
\colhead{(11)}
}
\startdata

DDO078		&	$0.09_{-0.00}^{+0.00}$	&	$148.3_{-2.0}^{+2.2}$	&	$0.06_{-0.00}^{+0.00}$	&	$159.9_{-3.9}^{+6.7}$	&	\nodata			&	\nodata			&	\nodata			&	\nodata			&	\nodata			&	\nodata			&	\nodata			&	\nodata			\\
IC4710		&	\nodata			&	\nodata			&	$0.24_{-0.08}^{+0.09}$	&	$50.2_{-0.4}^{+0.1}$	&	\nodata			&	\nodata			&	$0.24_{-0.02}^{+0.01}$	&	$37.1_{-1.7}^{+2.6}$	&	\nodata			&	\nodata			&	$0.3_{-0.02}^{+0.03}$	&	$28.6_{-2.6}^{+5.8}$	\\
NGC1258		&	$0.19_{-0.00}^{+0.01}$	&	$92.3_{-7.0}^{+1.3}$	&	$0.07_{-0.02}^{+0.02}$	&	$119.8_{-73.1}^{+61.9}$	&	$0.42_{-0.21}^{+0.43}$	&	$79.8_{-10.5}^{+17.9}$	&	\nodata			&	\nodata			&	\nodata			&	\nodata			&	\nodata			&	\nodata			\\
NGC3319		&	$0.29_{-0.01}^{+0.01}$	&	$83.3_{-7.6}^{+2.6}$	&	$0.08_{-0.00}^{+0.00}$	&	$75.3_{-5.8}^{+2.3}$	&	\nodata			&	\nodata			&	$0.04_{-0.00}^{+0.00}$	&	$74.4_{-3.0}^{+10.0}$	&	\nodata			&	\nodata			&	\nodata			&	\nodata			\\
NGC5334		&	$0.15_{-0.01}^{+0.01}$	&	$145.3_{-9.1}^{+10.2}$	&	$0.12_{-0.01}^{+0.01}$	&	$137.2_{-15.1}^{+8.8}$	&	$0.23_{-0.04}^{+0.04}$	&	$96.1_{-19.0}^{+12.0}$	&	$0.26_{-0.26}^{+0.02}$	&	$21.7_{-4.4}^{+7.2}$	&	\nodata			&	\nodata			&	\nodata			&	\nodata			\\

\nodata &\nodata &\nodata &\nodata &\nodata &\nodata &\nodata &\nodata &\nodata &\nodata &\nodata \\

\enddata

\vspace{-.5cm}
\tablecomments{For each filter, we list the ellipticity and position angle (in degrees measured North-to-East), derived from the best-fitting {\sc ishape} profile of the respective exposure, as listed in Table\,\ref{Table:reff}. 
}

\end{deluxetable}

\end{landscape}
\begin{landscape}
\begin{deluxetable}{llllllllllll}
\tabletypesize{\footnotesize}
%\rotate % For landscape mode
%\setlength{\tabcolsep}{0.15cm}
\tablecolumns{12}
\tablewidth{0pc} %%% <--- This is important!!! Otherwise it wont compile!!!!
\tablecaption{
Nuclear star cluster photometry. 
%See at the end the \tablecomments{}
(All 228 NSCs available in the online version of the table.)
\label{Table: F300-814W_mag}
}
\tablehead{

\colhead{OBJECT} &
\colhead{RA} &
\colhead{DEC} &
\colhead{$F606W_0$} &
\colhead{$F814W_0$} &
\colhead{$F450W_0$} &
\colhead{$F555W_0$} &
\colhead{$F675W_0$} &
\colhead{$F300W_0$} &
\colhead{$F336W_0$} &
\colhead{$F380W_0$} &
\colhead{$F439W_0$} \\
\colhead{} &
\colhead{hh:mm:ss} &
\colhead{dd:mm:ss} &
\colhead{mag} &
\colhead{mag} &
\colhead{mag} &
\colhead{mag} &
\colhead{mag} &
\colhead{mag} &
\colhead{mag} &
\colhead{mag} &
\colhead{mag} \\
\colhead{(1)} &
\colhead{(2)} &
\colhead{(3)} &
\colhead{(4)} &
\colhead{(5)} &
\colhead{(6)} &
\colhead{(7)} &
\colhead{(8)} &
\colhead{(9)} &
\colhead{(10)} &
\colhead{(11)} &
\colhead{(12)} 
}
\startdata

DDO078		&	10:26:27.14	&	67:39:10.18	&	$19.18\pm0.01_3$	&	$18.44\pm0.01_3$	&	\nodata			&	\nodata			&	\nodata			&	\nodata			&	\nodata			&	\nodata			&	\nodata			\\
IC4710		&	18:28:40.92	&	-66:59:09.63	&	\nodata			&	$18.26\pm0.01_3$	&	\nodata			&	$19.07\pm0.01_3$	&	\nodata			&	\nodata			&	\nodata			&	\nodata			&	$19.55\pm0.02_3$	\\
NGC1258		&	3:14:05.44	&	-21:46:27.95	&	$21.53\pm0.02_1$	&	$20.53\pm0.02_3$	&	$22.05\pm0.04_3$	&	\nodata			&	\nodata			&	\nodata			&	\nodata			&	\nodata			&	\nodata			\\
NGC3319		&	10:39:10.14	&	41:41:13.23	&	$18.96\pm0.01_2$	&	$18.55\pm0.01_4$	&	\nodata			&	$19.26\pm0.01_4$	&	\nodata			&	\nodata			&	\nodata			&	\nodata			&	\nodata			\\
NGC5334		&	13:52:54.68	&	-1:06:49.68	&	$20.01\pm0.01_3$	&	$19.28\pm0.02_3$	&	$20.85\pm0.02_3$	&	$20.35\pm0.02_3$	&	$19.77\pm0.02_3$	&	\nodata			&	\nodata			&	\nodata			&	\nodata			\\

\enddata

\vspace{-.5cm}

\tablecomments{Columns 1-3 list the name, RA, and DEC of the host galaxy. Columns 4-12 contain the CTE- and Galactic foreground reddening-corrected magnitudes of the NSCs in the WFPC2 photometric system. The subscripts indicate the detector used for the measurement.  
The applied Galactic foreground extinction is listed in Table\,\ref{Table:Galaxy sample}. 
}

\end{deluxetable}

\begin{deluxetable}{llllllllll}
\tabletypesize{\footnotesize}
%\rotate % For landscape mode
%\setlength{\tabcolsep}{0.15cm}
\tablecolumns{10}
\tablewidth{0pc} %%% <--- This is important!!! Otherwise it wont compile!!!!
\tablecaption{
Nuclear star cluster magnitudes in the Johnson/Cousins photometric system. \\
(All 228 NSCs are available in the online version of the table).
\label{Table: UBVI_mag}
}
\tablehead{

\colhead{OBJECT} &
\colhead{$V_0$} &
\colhead{$I_0$} &
\colhead{$B_0$} &
\colhead{$R_0$} &
\colhead{$U_0$} &
\colhead{$V_{\rm F555W}$} &
\colhead{$B_{\rm F439W}$} &
\colhead{$B_{\rm F380W}$} &
\colhead{$U_{\rm F300W}$} \\
\colhead{} &
\colhead{mag} &
\colhead{mag} &
\colhead{mag} &
\colhead{mag} &
\colhead{mag} &
\colhead{mag} &
\colhead{mag} &
\colhead{mag} &
\colhead{mag} \\
\colhead{(1)} &
\colhead{(2)} &
\colhead{(3)} &
\colhead{(4)} &
\colhead{(5)} &
\colhead{(6)} &
\colhead{(7)} &
\colhead{(8)} &
\colhead{(9)} &
\colhead{(10)}
}
\startdata

DDO078		&	$19.4\pm0.01$	&	$18.28\pm0.01$	&	\nodata		&	\nodata		&	\nodata		&	\nodata	&	\nodata	&	\nodata	&	\nodata	\\
IC4710		&	$18.96\pm0.01$	&	$18.1\pm0.01$	&	$19.4\pm0.02$	&	\nodata		&	\nodata		&	$18.96$	&	$19.4$	&	\nodata	&	\nodata	\\
NGC1258		&	$21.92\pm0.02$	&	$20.36\pm0.02$	&	$22.0\pm0.05$	&	\nodata		&	\nodata		&	\nodata	&	\nodata	&	\nodata	&	\nodata	\\
NGC3319		&	$19.98\pm0.01$	&	$18.4\pm0.01$	&	\nodata		&	\nodata		&	\nodata		&	$19.15$	&	\nodata	&	\nodata	&	\nodata	\\
NGC5334		&	$20.23\pm0.01$	&	$19.08\pm0.02$	&	$20.8\pm0.02$	&	$20.69\pm0.02$	&	\nodata		&	$20.24$	&	\nodata	&	\nodata	&	\nodata	\\

\nodata &\nodata &\nodata &\nodata &\nodata &\nodata &\nodata &\nodata &\nodata &\nodata \\

\enddata

\vspace{-.5cm}
\tablecomments{The Johnson/Cousins magnitudes are calculated from the HST/WFPC2 magnitudes as described in \S\,\ref{Sect: NSC photometry}. Columns (2)-(6) contain the magnitudes derived from the filters F606W, F814W, F450W, F675W, or F336W. 
If those are unavailable, we adopt measurements from alternative filters listed in columns (7)-(10).
}

\end{deluxetable}

\end{landscape}

\bibliographystyle{aa}
\bibliography{references}

\appendix

\section[]{Additional figures}\label{Appendix}

\begin{figure*}
\centering
\subfloat[]{\includegraphics[page=1,width=0.5\textwidth]{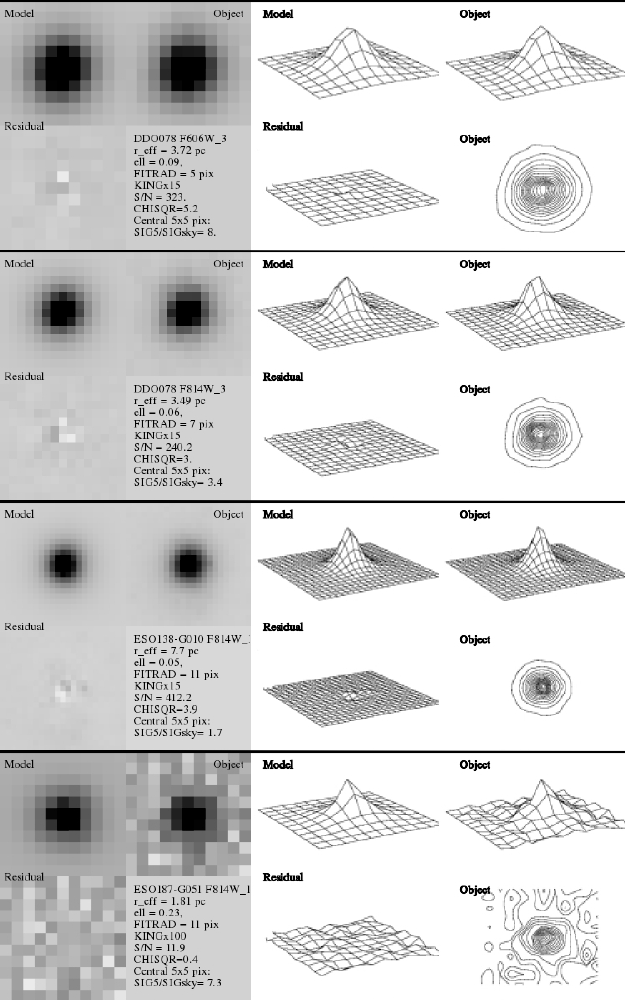}}\hspace*{0.2cm}
\subfloat[]{\includegraphics[page=2,width=0.5\textwidth]{Fig_A1.pdf}}
\caption{Images and surface/contour plots of the 228 NSCs, their best 
fit models, and fit residuals (data - model). A summary of the fit statistics is also provided. The full figure is available online.
}\label{fig:Residuals map}
\end{figure*}

\label{lastpage}

\end{document}